\documentclass[12pt,natbib,iop,hyperref]{aa}
\usepackage{natbib}
\pdfoutput=1
\bibpunct{(}{)}{;}{a}{}{,}
\usepackage{multirow}
\usepackage{color}
\usepackage[varg]{txfonts}

\begin{document}

\newcommand{\kms}{\>{\rm km}~\,{\rm s}^{-1}}
\newcommand{\reff}{r_{\rm{eff}}}
\newcommand{\msol}{M$_{\odot}$}
\newcommand{\zsol}{Z_{\odot}}
\newcommand{\inverse}[1]{{#1}^{-1}}
\newcommand{\invvar}{\inverse{C}}
\newcommand{\dd}{{\rm d}}
\newcommand{\cgs}{erg s$^{-1}$ cm$^{-2}$}
\newcommand{\pf}{\texttt{Platefit} }
\newcommand{\magphys}{\texttt{MAGPHYS} }
\newcommand{\udft}{\textsf{udf-10~}}
\newcommand{\udftns}{\textsf{udf-10}}

\title{The MUSE Hubble Ultra Deep Field Survey:}
\subtitle{IV. Global properties of C III] emitters\thanks{Based on observations made with ESO telescopes at the La Silla Paranal Observatory under program IDs 60.A-9100(C), 094.A-2089(B), 095.A-0010(A), 096.A-0045(A), and 096.A-0045(B).  This work is also based on observations made with the NASA/ESA Hubble Space Telescope, programs GO-12099 and 12177, obtained at the Space Telescope Science Institute, which is operated by the Association of Universities for Research in Astronomy, Inc., under NASA contract NAS 5-26555.}}
\author{Michael V. Maseda\inst{\ref{inst1}}\thanks{NOVA Fellow; \email{maseda@strw.leidenuniv.nl}}
\and Jarle Brinchmann\inst{\ref{inst1},\ref{instp}}
\and Marijn Franx\inst{\ref{inst1}}
\and Roland Bacon\inst{\ref{inst2}}
\and Rychard J. Bouwens\inst{\ref{inst1}}
\and Kasper B. Schmidt\inst{\ref{inst3}}
\and Leindert A. Boogaard\inst{\ref{inst1}}
\and Thierry Contini\inst{\ref{inst5},\ref{inst6}}
\and Anna Feltre\inst{\ref{inst2}}
\and Hanae Inami\inst{\ref{inst2}}
\and Wolfram Kollatschny\inst{\ref{inst8}}
\and Raffaella A. Marino\inst{\ref{inst7}}
\and Johan Richard\inst{\ref{inst2}}
\and Anne Verhamme\inst{\ref{inst2},\ref{inst4}}
\and Lutz Wisotzki\inst{\ref{inst3}}}

\institute{Leiden Observatory, Leiden University, P.O. Box 9513, 2300 RA, Leiden, The Netherlands\label{inst1}
\and Instituto de Astrof{\'\i}sica e Ci{\^e}ncias do Espa\c{c}o, Universidade do Porto, CAUP, Rua das Estrelas, PT4150-762 Porto, Portugal\label{instp}
 \and Univ Lyon, Univ Lyon1, Ens de Lyon, CNRS, Centre de Recherche Astrophysique de Lyon UMR5574, F-69230, Saint-Genis-Laval, France\label{inst2}
 \and Leibniz-Institut f\"ur Astrophysik Potsdam (AIP), An der Sternwarte 16, 14482 Potsdam, Germany\label{inst3}
  \and IRAP, Institut de Recherche en Astrophysique et Plan\'etologie, CNRS, 14 avenue \'Edouard Belin, 31400 Toulouse, France\label{inst5}
   \and Universit\'e de Toulouse, UPS-OMP, 31400 Toulouse, France\label{inst6}
        \and Institut f\"ur Astrophysik, Universit\"at G\"ottingen, Friedrich-Hund-Platz 1, 37077 G\"ottingen, Germany\label{inst8}
    \and ETH Z\"urich, Institute of Astronomy,Wolfgang-Pauli-Str. 27, 8093 Z\"urich, Switzerland\label{inst7}
      \and Observatoire de Gen\`eve, Universit\'e de Gen\`eve, 51 Ch. des Maillettes, 1290 Versoix, Switzerland\label{inst4}}

\titlerunning{Global properties of C III] emitters}
\authorrunning{M. V. Maseda et al.}
\date{June 27, 2017}

\abstract{The C III] $\lambda\lambda$1907,1909 emission doublet has been proposed as an alternative to Lyman-$\alpha$ in redshift confirmations of galaxies at $z \gtrsim 6$ since it is not attenuated by the largely neutral intergalactic medium at these redshifts and is believed to be strong in the young, vigorously star-forming galaxies present at these early cosmic times.  We present a statistical sample of 17 C III]-emitting galaxies beyond $z \sim 1.5$ using $\sim$30 hour deep VLT/MUSE integral field spectroscopy covering 2 square arcminutes in the \textit{Hubble} Deep Field South (HDFS) and Ultra Deep Field (UDF), achieving C III] sensitivities of $\sim 2~ \times$ 10$^{-17}$ erg s$^{-1}$ cm$^{-2}$ in the HDFS and $\sim 7~ \times$ 10$^{-18}$ erg s$^{-1}$ cm$^{-2}$ in the UDF.  The rest-frame equivalent widths range from 2 to 19 \AA.  These 17 galaxies represent $\sim$3\% of the total sample of galaxies found between $1.5 \lesssim z \lesssim 4$.  They also show elevated star formation rates, lower dust attenuation, and younger mass-weighted ages than the general population of galaxies at the same redshifts.  Combined with deep slitless grism spectroscopy from the HST/WFC3 in the UDF, we can tie the rest-frame ultraviolet C III] emission to rest-frame optical emission lines, namely [O III] $\lambda$5007, finding a strong correlation between the two.  Down to the flux limits that we observe ($\sim 1~ \times$ 10$^{-18}$ erg s$^{-1}$ cm$^{-2}$ with the grism data in the UDF), all objects with a rest-frame [O III] $\lambda\lambda$4959,5007 equivalent width in excess of 250 \AA, the so-called Extreme Emission Line Galaxies, have detections of C III] in our MUSE data.  More detailed studies of the C III]-emitting population at these intermediate redshifts will be crucial to understand the physical conditions in galaxies at early cosmic times and to determine the utility of C III] as a redshift tracer.}


\maketitle

\section{INTRODUCTION}
Large samples of candidate $z>6$ galaxies have been constructed with the Lyman-break technique on deep imaging data, using Hubble Space Telescope (HST) Advanced Camera for Surveys (ACS) data at optical wavelengths \citep{2003MNRAS.342..439S,2004ApJ...600L..99D} before moving to higher redshifts and larger samples with HST Near Infrared Camera and Multi-Object Spectrometer (NICMOS) data at near-infrared wavelengths \cite[e.g.,][]{2004ApJ...616L..79B}.  Eventually, the installation of the Wide Field Camera 3 (WFC3) on the HST, with a higher sensitivity and larger field of view than NICMOS, has led to samples of hundreds of robust photometric candidates \cite[e.g.,][]{2010MNRAS.409..855B,2011ApJ...727L..39T,2013ApJ...763L...7E,2013MNRAS.432.2696M,2013ApJ...773...75O,2013ApJ...768..196S,2014ApJ...786...57S,2015ApJ...803...34B,2015MNRAS.452.1817B,2015ApJ...810...71F}.  Complementary results have also been obtained from larger and often shallower ground-based imaging campaigns \cite[e.g.,][]{2005PASJ...57..165T,2009ApJ...706.1136O,2010ApJ...723..869O,2010ApJ...725..394H,2012ApJ...752..114S,2013ApJ...768...56T,2014ApJ...797...16K,2015MNRAS.451..400M}.  While the number of photometric candidates is impressive, only small subsamples of these candidates have been confirmed spectroscopically with Lyman-$\alpha$ \cite[e.g.,][]{2010ApJ...725..394H,2010ApJ...723..869O,2011ApJ...734..119K,2012ApJ...761...85K,2011ApJ...743..132P,2011ApJ...730L..35V,2012ApJ...752..114S,2013Natur.502..524F,2013ApJ...778..102O,2015ApJ...804L..30O,2015MNRAS.454.1393S,2015ApJ...808..139S,2015ApJ...810L..12Z,2016ApJ...818...38S,2016ApJ...826..113S,2016ApJ...827L..14T,2017ApJ...837L..21L}.  See \citet{2016ARAA..54..761S} and references therein for a comprehensive review.

This is not for lack of trying, as many studies have returned negative results in the search for the highest redshift Lyman-$\alpha$ emitters \cite[e.g.,][]{2010ApJ...725L.205F,2012MNRAS.427.3055C,2013MNRAS.430.3314B,2013ApJ...775L..29T,2014MNRAS.440.2375M,2014ApJ...793..113P}.  Emission line redshifts are often preferred compared to the direct spectral detection of the continuum break (the so-called Lyman break) due to the faint continuum levels in high-redshift galaxies, although a handful of exceptional cases exist where a continuum break is detected at high redshift \cite[e.g.,][]{2009Natur.461.1254T,2016ApJ...819..129O}.  Even a combined 52-hour VLT/FORS2 spectrum of one of the brightest and most robust $z\sim7$ candidates in the \textit{Hubble} Ultra Deep Field (UDF) does not show Lyman-$\alpha$ or continuum emission in the expected wavelength range \citep{2014AA...569A..78V}.  Much of the difficulty is believed to be caused by the increasingly neutral intergalactic medium at these redshifts \cite[e.g.,][]{2007MNRAS.379..253D}, which would attenuate the Lyman-$\alpha$ emission that is relied upon to confirm the redshifts.

Some authors have suggested that other, relatively strong emission lines can be used to confirm redshifts at $z\gtrsim6$, namely the semi-forbidden C III] doublet at (vacuum) 1906.7 and 1908.7 \AA$~$\citep{2014MNRAS.445.3200S,2015ApJ...805L...7Z}.  The actual samples of galaxies where this doublet is observed has remained, until recently \citep{2016arXiv161206866D}, very small and is predominately limited to $z\sim0$ galaxies \citep{2011AJ....141...37L,2015ApJ...814L...6R} and blue, low-mass $z\sim1-3$ galaxies \citep{2010ApJ...719.1168E,2017arXiv170104416A} of which the majority are strongly gravitationally lensed \citep{2010MNRAS.406.2616C,2012MNRAS.427.1973C,2014ApJ...790..144B,2014MNRAS.445.3200S,2015ApJ...814L...6R,2016MNRAS.456.4191P}.  New results at $z \sim 6-7$ from \citet{2015MNRAS.450.1846S,2017MNRAS.464..469S} have shown strong C III] emission and posit that the cause is an extremely hard ionization field produced by low-metallicity stellar populations.

However, it must be stressed that the samples of C III] emission in star-forming galaxies are currently small and, at least at $z > 0$, biased toward the lowest mass, bluest galaxies.  While these galaxies may be similar to galaxies forming at the earliest cosmic times, they are not fully representative of the general galaxy population; \citet{2003ApJ...588...65S} found C III] at approximately 10\% of the flux of Lyman-$\alpha$ in a stacked spectrum of $\sim$ 1000 Lyman Break Galaxies (LBGs) at $z\sim3$ \cite[cf.][who show that this flux ratio varies between samples at fixed redshift]{2015ApJ...814L...6R}.

The general picture of C III], then, is that it is nearly omnipresent in star-forming galaxies at $z > 0$ albeit at relatively faint fluxes with a strength that may increase with a higher ionization parameter and/or decreasing stellar mass or gas phase metallicity.  Shocks and active galactic nuclei (AGN) are also capable of producing C III] emission in galaxies.  The relative contributions of these mechanisms compared to star formation is still unknown.  Given this, there is a clear need for a larger and more representative sample of galaxy spectroscopy that is capable of finding C III] even at modestly low fluxes.  Here we combine extremely deep optical spectroscopy from MUSE \cite[the Multi-Unit Spectroscopic Explorer;][]{2010SPIE.7735E..08B} on the \textit{ESO} Very Large Telescope (VLT) with deep multi-band photometry in the UDF and the \textit{Hubble} Deep Field South (HDFS) along with HST/WFC3 near-infrared slitless grism spectroscopy (in the UDF only) to systematically obtain a sample of (unlensed) $1.5 \lesssim z \lesssim 4$ C III] emitters.  Given the amount of ancillary information present in these areas, we can investigate the prevalence of C III] emission with properties such as stellar mass, (specific) star formation rate, and UV luminosity.

We refer to the combined [C III] $\lambda$1907 and C III] $\lambda$1909 doublet as C III] throughout except when otherwise noted.  We adopt a flat $\Lambda$CDM cosmology ($\Omega_m=0.3$, $\Omega_\Lambda=0.7$, and H$_0=70 ~$km s$^{-1}$ Mpc$^{-1}$) and AB magnitudes \citep{1974ApJS...27...21O}.
\section{Data}

\subsection{MUSE Observations}
\label{sec:z}
We used the \udftns, which is the deepest pointing in a larger 3$'~\times~$3$'$ mosaic of nine MUSE pointings in the UDF.  Details of the reduction of the \udft data are given in \citet{Bacon2017}, hereafter referred to as Paper I. In this reduction, 156 individual exposures were combined into a single 1$'$x1$'$ MUSE cube with a median exposure time per pixel of 31.6 hours.  The effective FWHM of the seeing (white light) is 0.65$''$. 

Because of its similar depth ($\sim$27 hours), we also incorporated the HDFS data taken as part of the commissioning of MUSE \citep{2015AA...575A..75B}.  We used a new post-processing of the MUSE data cube, v1.35 \footnote{\url{data.muse-vlt.eu/HDFS/v1.30/DATACUBE-HDFS-1.34.fits.gz}}, similar to that presented in \citet{2016arXiv160501422B}.  The main differences between this reduction and the publicly released cube, which used the pipeline described in \citet{2012SPIE.8451E..0BW}, is improved flat fielding and sky subtraction (using CubeEx; Cantalupo, in prep.).  The average FWHM of the seeing (white light) in the HDFS cube is 0.77$''$.

\begin{figure*}
\begin{center}
\includegraphics[height=.45\textheight]{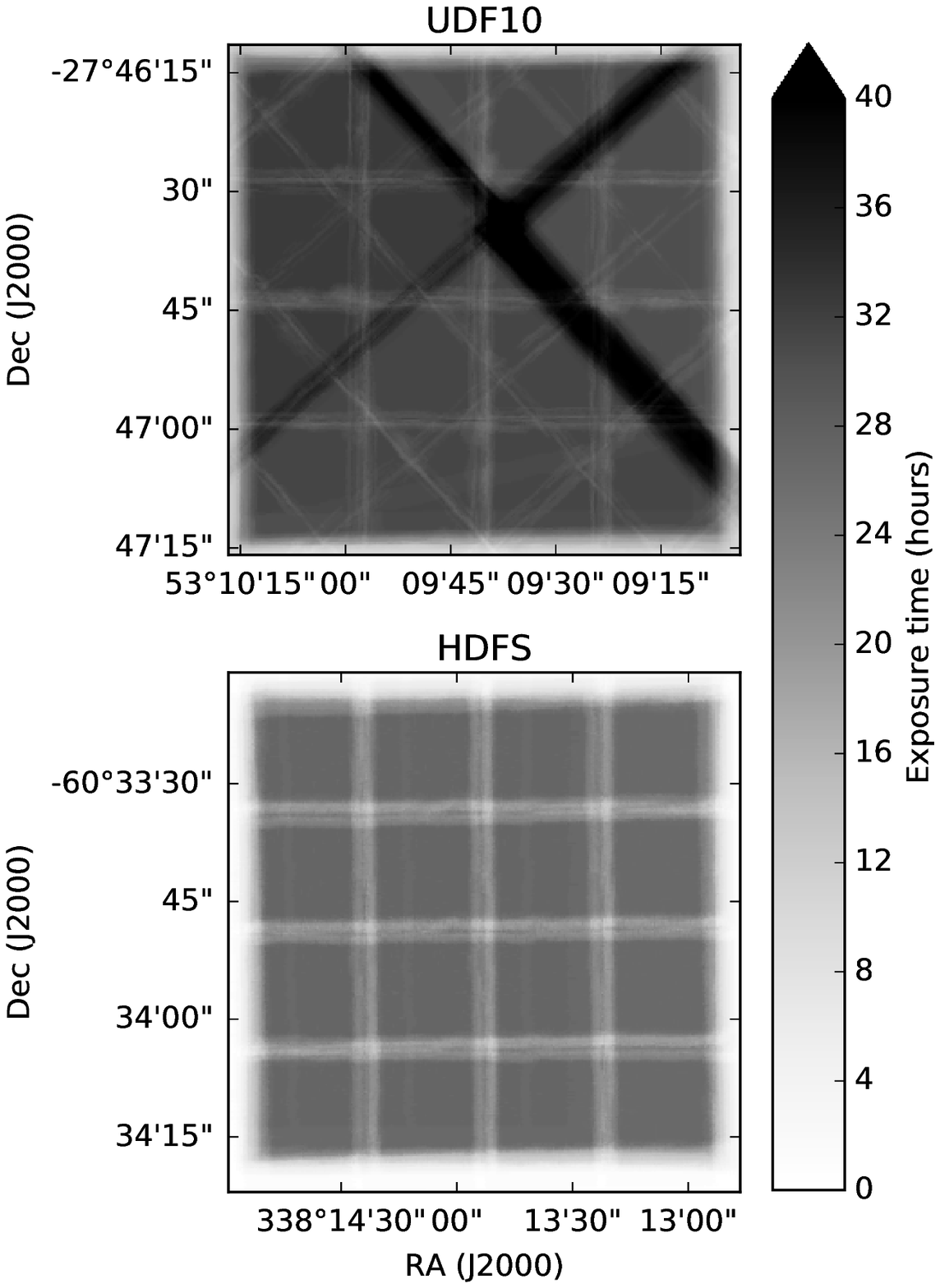}
\includegraphics[height=.45\textheight]{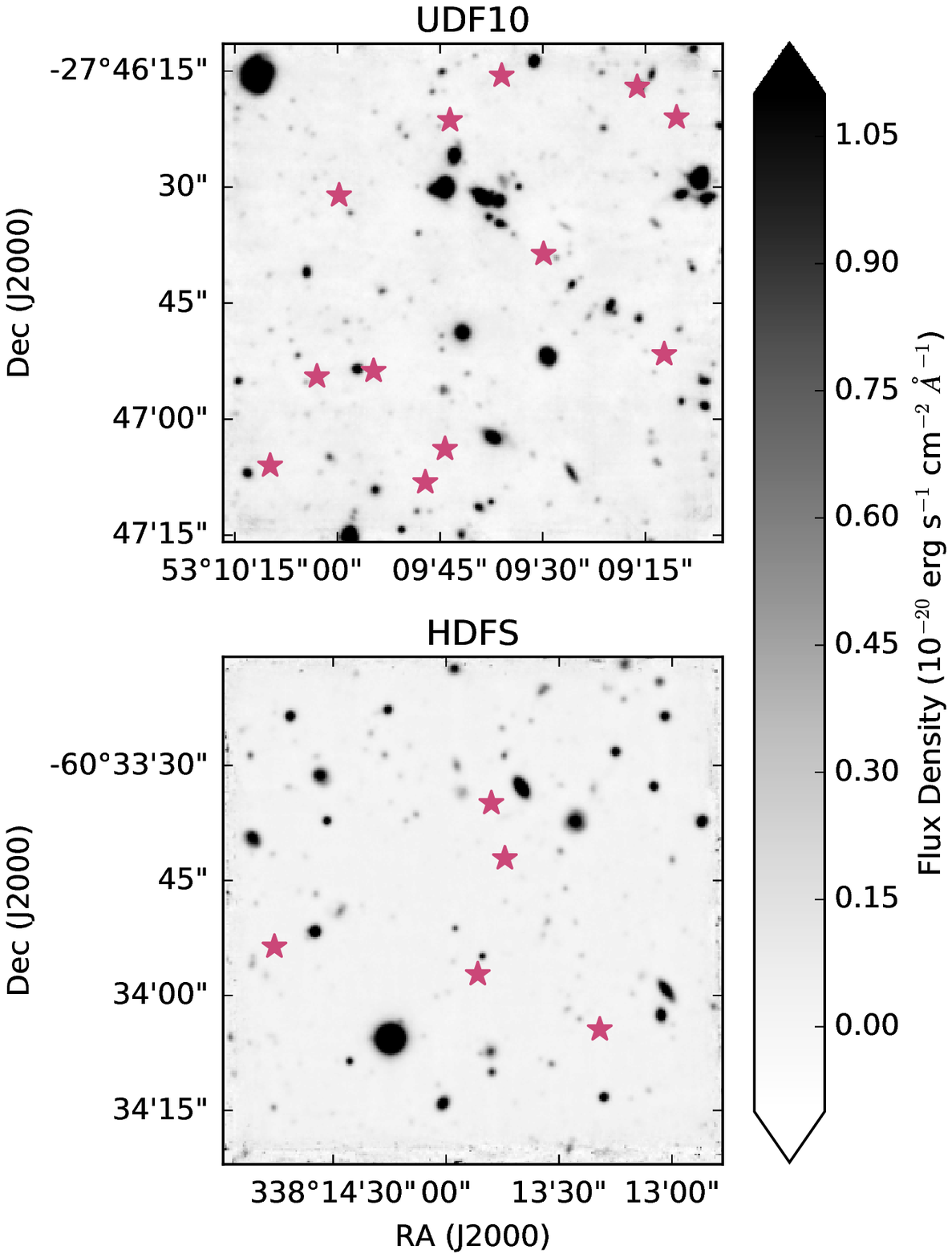}
\end{center}
\caption{Exposure maps (left) and white light images (right) for the two MUSE data cubes used here: \udft version 0.42 (top) and HDFS version 1.34 (bottom).  The pink stars in the white light images show the positions of C III] emitters (see Section \ref{sec:ciii} and Table \ref{tab:ciii}).}
\label{fig:exps}
\end{figure*}

White light images (the MUSE cube flattened along the spectral direction)  and exposure maps for the two fields are shown in Figure \ref{fig:exps}.

%\subsubsection{Emission line redshifts and identifications}
For the remainder of this analysis, we discuss one-dimensional spectra extracted from the MUSE data cubes. For sources that are unresolved at the spatial resolution of MUSE, a weighted extraction using a single kernel is insufficient since, owing to the wavelength-dependent point spread function (PSF), we would be adding too little flux in the blue and relatively more in the red.  Thus we integrated the source flux in a region defined by its original segmentation area convolved with a Gaussian of 0.6$''$ FWHM to take into account the MUSE resolution, weighted by the wavelength-dependent PSF in the spectrum extraction.  We centered the PSF on the center of the object when performing the summation.  As noted in Paper I, this is the optimal way of extracting the spectrum for small and/or faint objects.

Details of the redshift determinations for the HDFS data are given in \citet{2015AA...575A..75B} and for the \udft data in \citeauthor{Inami2017} (\citeyear{Inami2017}; hereafter Paper II).  We briefly summarize the general methods here.

The starting samples for redshift determinations are the respective photometric catalogs for the field.  For continuum-selected objects, one-dimensional spectra were extracted from the MUSE cubes.  We compared the spectra to spectroscopic templates derived from MUSE data for an initial redshift determination and then were inspected  by multiple individuals for confirmation.  The majority of redshifts come from emission lines with the additional constraint that the two-dimensional profile of the emission line in a pseudo-narrowband image should be coherent.  We can only accurately determine redshifts
from absorption features in bright sources, where the continuum is resolved well. 

As noted in \citet{2015AA...575A..75B}, some objects have emission lines with high equivalent widths (EWs) in excess of 100 \AA\ that are easily detectable with MUSE but the galaxy is otherwise too faint for a continuum detection, even in the deep HST imaging that exists in these fields.  The tool, 
 \texttt{ORIGIN}, has been developed to search for these emission lines in the full MUSE data cubes \cite[based on algorithms described in Paper I;][]{Bourguignon201232, paris:tel-00933827}; this tool accounts for a spectrally varying instrumental PSF and the spectral similarities of neighboring pixels.  This algorithm is applied to the \udft data.  For reference, this method detects 152 plausible emission line sources in the \udft that were missed in the input photometric catalog \citep{2015AJ....150...31R}, but have a counterpart in the HST imaging from which photometry can be extracted; 32 of these sources have no HST counterpart at all.  Another algorithm, \texttt{MUSELET}\footnote{\url{http://mpdaf.readthedocs.io/en/latest/muselet.html}}, applied to the HDFS data, detects emission lines in pseudo-narrowband images following an emission line profile with a velocity of $\sigma = 100 \kms$ \citep{2015AA...575A..75B,2015MNRAS.446L..16R}.  One-dimensional spectra were extracted for these sources and redshifts were determined in an identical way to the photometrically selected sources.

In total, there are 308 redshift determinations in the \udft cube and in this version of the HDFS cube there are 239 redshifts (all confidence levels; Paper II).  We used these redshifts as inputs to a C III] line search as described in Section \ref{sec:lines}.

\subsection{Photometry}
\label{sec:phot}

In the HDFS, we used the catalogs from \citet{2007ApJ...655...51W}, who combined the optical through near-IR imaging from HST/WFPC2 \cite[$F300W, F450W, F606W,$ and $F814W$;][]{2000AJ....120.2747C} and VLT/ISAAC \cite[$J_s,~H,$ and $K_s$;][]{2003AJ....125.1107L} with \textit{Spitzer Space Telescope} IRAC imaging in the 3.6, 4.5, 5.8, and 8 $\mu$m bands. In the \udftns, the MUSE redshift identifications were performed using the \citet{2015AJ....150...31R} photometric catalog as an input.  This catalog contains 11 \textit{HST}/WFC3 and ACS photometric bands, but does not include photometric coverage redward of 1.7 $\mu$m.  Therefore, we chose to use the 26-band 3D-HST photometric catalog of \citet{2014ApJS..214...24S}, which includes both ground-based and HST-based optical/near-IR photometry and IRAC photometry, supplemented by 3D-HST grism spectroscopic data and \textit{Spitzer}/MIPS 24 $\mu$m photometry from \citet{2016ApJS..225...27M}.  We refer to this combined photometric catalog as the 3D-HST photometry in the following.  This catalog provides grism redshifts that are a crucial addition to our spectral energy distribution (SED) fitting in Section \ref{sec:sed} in cases where we do not have a redshift from MUSE.  The near-IR coverage provided is also important since, for example, \citet{2007ApJ...655...51W} demonstrated the constraining power that observations at \textit{Spitzer}/IRAC wavelengths provide on the derived SED-fitting parameters.  Finally, the 3D-HST photometric bands cover a similar wavelength range as those used in the HDFS, making the derived results comparable between the fields.

For all galaxies we calculated $\beta$, the UV continuum slope, by performing a power-law fit to all photometric data points where the central wavelength of the filter covers the rest-frame wavelength range $1300 < \lambda/$\AA$~< 2500$ for the galaxy, requiring at least two photometric points in this range.  For non stellar objects that do not have a MUSE redshift, we computed photometric redshifts with \texttt{EAZY}\footnote{\url{https://github.com/gbrammer/eazy-photoz/}} \citep{2008ApJ...686.1503B}.  Within \texttt{EAZY}, we adopted an $R$-band ($F606W$) prior, which defines the prior probability distribution on redshift for a given apparent magnitude, $p(z|m_0)$.  In addition to the standard five templates, we included a young, dusty template and an old, red galaxy template \cite[as described in][]{2011ApJ...735...86W}.

\subsection{HST grism spectroscopy}
\label{sec:grismdat}
The grism spectra here come from deep stacks of all available G141 grism data in the \udft region, which include eight orbits from the 3D-HST program \cite[GO-12177;][]{2012ApJS..200...13B,2016ApJS..225...27M} and nine orbits from supernova follow-up observations from the CANDELS program \cite[GO-12099;][]{2012ApJ...746....5R}. The grism observations are combined in a similar way to that mentioned in \citet{2013ApJ...765L...2B} and \citet{2013arXiv1305.2140V}, but specifically using the \texttt{Grizli}\footnote{\url{https://github.com/gbrammer/grizli/}} code.  Line fluxes are determined according to the spatial profile of the galaxy from its F140W morphology, which is similar in wavelength coverage to the G141 grism; cf. the PSF-weighting used in the MUSE spectral extractions: in the case of a compact object, like the majority of our C III] emitters, these flux determinations should be comparable.  In our calculations of equivalent widths we determine the continuum level from the broadband photometry (either F125W or F160W, depending on the wavelength of the line), correcting for the contribution of the emission lines; in most cases the continuum is not detected spectroscopically.

\section{Methods}

\subsection{Emission line recovery}
\label{sec:lines}
As described in Section \ref{sec:z}, redshifts for MUSE detections were determined using a combination of manual inspections and automated template fitting.  For each object with a redshift, we fit the spectrum with \pf \citep{2004ApJ...613..898T, 2004MNRAS.351.1151B}, which constrains the local continuum level and measures the strengths of the emission and absorption features in the observed wavelength range.  \pf allows for velocity shifts up to 300 $\kms$ of all lines from the input ``systemic'' redshift. In cases where Ly-$\alpha$ is significantly offset from the true systemic redshift of the galaxy, as traced by C III], then \pf would not correctly recover C III] or other spectral features.  Additionally, since the input redshifts come from template matches they can be off by $\sim 0.1$ \AA$~$ owing to, for example, variable line strengths or widths, so this feature further refines the redshift determination.  

In order to assess the practical flux limits of the (one-dimensional) MUSE spectra, we inserted an artificial emission line doublet with the same rest-frame spacing as the C III] doublet, centered at rest-frame 2000 \AA$~$according to the redshifts described in Section \ref{sec:z}.  The doublet has a fixed 1907/1909 flux ratio of 1.53 \cite[the low-density limit from][but cf. Section \ref{sec:ciiidens}]{1992ApJ...389..443K} and a fixed width of 80 $\kms$ ($\sigma$, which is the default \pf line width).  The position of the artificial 2000 \AA$~$doublet does not overlap with any other major nebular or stellar spectral feature and is close enough to the actual position of C III] such that the continuum determination (plus the associated uncertainties) and the effect of bright OH skylines at the highest redshifts, which are most prevalent in the red, are similar.  

%get these results from pydoublet_results_combined.py
\begin{table}
\caption{Limiting emission doublet sensitivities}              % title of Table
\label{tab:sens}      % is used to refer this table in the text
\centering                                      % used for centering table
\begin{tabular}{l|ccc}          
\hline\hline  

Detection & 90\% & 75\% & 50\% \\ threshold &Recovery&Recovery&Recovery\\
\hline
& & & \\
\textbf{HDFS} & & &\\
\hline
3-$\sigma$ & -16.70 & -16.89 & -17.09\\
 & & & \\
\textbf{\udft} & & &\\
\hline
3-$\sigma$ & -17.13 & -17.32 & -17.52\\
\end{tabular}
\tablefoot{All fluxes are in log cgs units (erg s$^{-1}$ cm$^{-2}$).  These values are extracted from the data shown in Figure \ref{fig:recovery}.  The detection thresholds quoted here use the integrated doublet S/N values from \texttt{Platefit}, and are normalized at an input flux of 10$^{-16}$ erg s$^{-1}$ cm$^{-2}$.  For the remainder of this work, we adopt 3$\sigma$ as the detection threshold.}
\end{table}

\begin{figure}
\includegraphics[width=.45\textwidth]{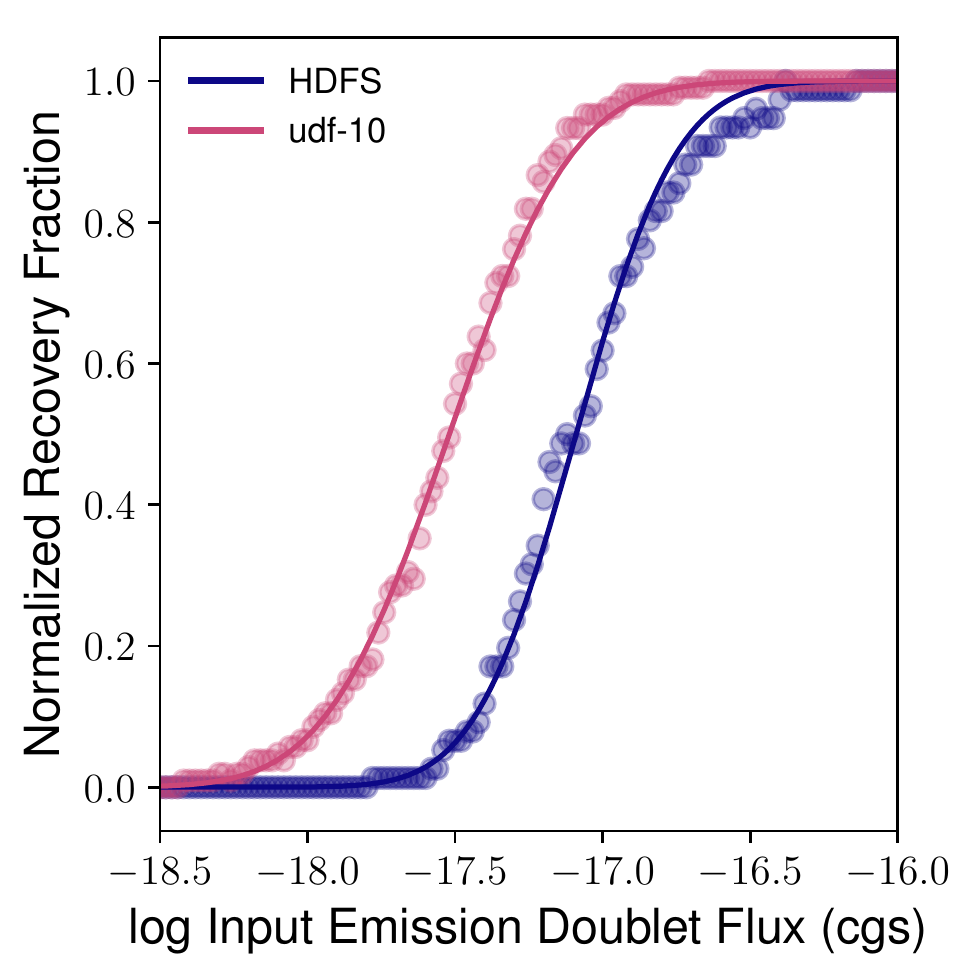}
\caption{Results from the test inserting a simulated emission line doublet at 2000 \AA $~$(rest frame) and recovering the line using \pf at a significance of 3$\sigma$.  The curves show fits for the lognormal cumulative distribution functions of the data assuming Poisson errors in the number of recovered objects.  Numerical results for the flux limits at fixed completeness levels (90\%, 75\%, and 50\%, normalized at an input line flux of 10$^{-16}$ erg s$^{-1}$ cm$^{-2}$) and using fixed detection thresholds (3$\sigma$ in integrated doublet S/N) are shown in Table \ref{tab:sens}.}
\label{fig:recovery}
\end{figure}

The results of this exercise are shown in Figure \ref{fig:recovery} and Table \ref{tab:sens}.  The normalized recovery fraction is the fraction of the input lines that could be retrieved successfully by \pf at a line flux of 10$^{-16}$ erg s$^{-1}$ cm$^{-2}$.  At this flux level, the recovery curves for both fields are flat.  In approximately 14\% of the spectra tested here (30 out of 211), no artificial line could be recovered ($>$ 3$\sigma$) at this flux level.  This is primarily due to severe skyline contamination in the MUSE optical spectra, particularly at red wavelengths.  We can therefore consider the results to be the wavelength-averaged recovery fraction for lines that fall within clean wavelength windows since the MUSE line sensitivity is, modulo severe skyline contamination, relatively constant with wavelength (see Paper I).

Even though we could observe C III] in MUSE up to a redshift of $\sim$ 4, the highest redshift C III] emitter in our sample (described in Section \ref{sec:ciii}) is at z $=2.9$.  We attribute this to (1) the difficulty in having a clear line identification and recovery at wavelengths longer than $\sim$7500 \AA, where OH skylines become stronger in the MUSE spectra and (2) the larger flux uncertainties associated with these wavelengths due to the skylines.  While we present line flux sensitivities in Table \ref{tab:sens} and Figure \ref{fig:recovery} that are averaged over all wavelengths, in reality the probability that a real emission line is completely or partially masked by a skyline is a function of redshift.  As described above, the number of clean wavelength windows in which we could recover an emission feature at an arbitrary flux level (i.e., 10$^{-16}$ erg s$^{-1}$ cm$^{-2}$, as in Figure \ref{fig:recovery} and Table \ref{tab:sens}) decreases with wavelength and hence redshift.  For the brightest fluxes measured in Section \ref{sec:ciii}, 7$\times$10$^{-18}$ erg s$^{-1}$ cm$^{-2}$, we expect the \udft data to be $\sim$90\% complete when averaged over all clean wavelengths.  For the faintest, 5$\times$10$^{-19}$ erg s$^{-1}$ cm$^{-2}$, we are only $\sim$5\% complete when averaged over all clean wavelengths.

The determination of a line equivalent width is dependent on a line flux measurement and a continuum measurement.  Particularly in the faintest sources, accurate determinations of the continuum level can be difficult to make.  In order to explore the combined uncertainties between the line and continuum fits, we performed a Monte Carlo simulation for each spectrum by creating a series of mock spectra where each flux point is randomly perturbed from its measured value according to the statistical variance at that point.  These perturbations are Gaussian and the variances include an empirical correction for the correlation between pixels (see Section 3.1.5 of Paper I for details). We then run \pf on the mock spectra and determine the C III] equivalent width.  This procedure is repeated 500 times on each object.  

We therefore determined the (rest-frame) C III] equivalent width according to
\begin{equation}
EW_{C III],0} = mean \left( \frac{F_{1907,i}+F_{1909,i}}{0.5\times(Cont_{1907,i}+Cont_{1909,i})} \right),
\end{equation}
where $F_i$ and $Cont_i$ refer to the line flux and continuum in the $i^{th}$ Monte Carlo simulation determined with \pf for each component of the doublet; quoted uncertainties in EW, which are listed in Table \ref{tab:ciii}, are the 1$\sigma$ standard deviations of these same distributions.  This value of EW is defined to be positive for emission lines.  The means and standard deviations of the flux measurements from these simulations are also listed in Table \ref{tab:ciii}.

Our criteria for a C III] detection is a combined S/N in the doublet of 3 (i.e., the yellow curves in Figure \ref{fig:recovery}), a positive measured flux value in both 1907 and 1909 \AA$~$components, a velocity width $\sigma < 200 \kms$ in each component, and a combined doublet rest-frame equivalent width greater than 1 \AA.  The constraint on the velocity width removes cases where large-scale continuum features are fit as emission lines, and the constraint on the equivalent width compensates for the flux-limited nature of our survey.  

\subsection{Spectral energy distribution fitting}
\label{sec:sedfit}
We used \magphys \citep{2008MNRAS.388.1595D} with the high-z extension \citep{2015ApJ...806..110D}, which includes new star formation histories and new dust priors, to fit the broadband SEDs of the galaxies.  By default the \magphys high-z extension only allows for a minimum stellar mass of $10^8$ \msol, but we modified this limit to 10$^6$ \msol $ $ to account for the depth of the broadband imaging in these fields.

There are 777 (3D-HST) photometric sources in the MUSE \udft footprint and 544 photometric sources in the MUSE HDFS footprint.  Redshift determinations for each photometric source are made using, in order of reliability, (1) MUSE optical spectroscopy, (2) other ground-based spectroscopy from the literature \cite[see discussion in][]{2014ApJS..214...24S}, (3) WFC3/G141 grism spectroscopy (\udft only), and (4) photometric redshifts using \texttt{EAZY} \citep{2008ApJ...686.1503B}.  Applying a cut in redshift where we would be able to observe C III] with MUSE (1.49 $< z <$ 3.90) yields 322 sources in the \udft and 331 sources sources in the HDFS.  This sample of 653 galaxies is referred to as the total photometric sample throughout.

\section{ C III] detections}
\label{sec:ciii}

\begin{figure}
\begin{center}
\includegraphics[width=.5\textwidth]{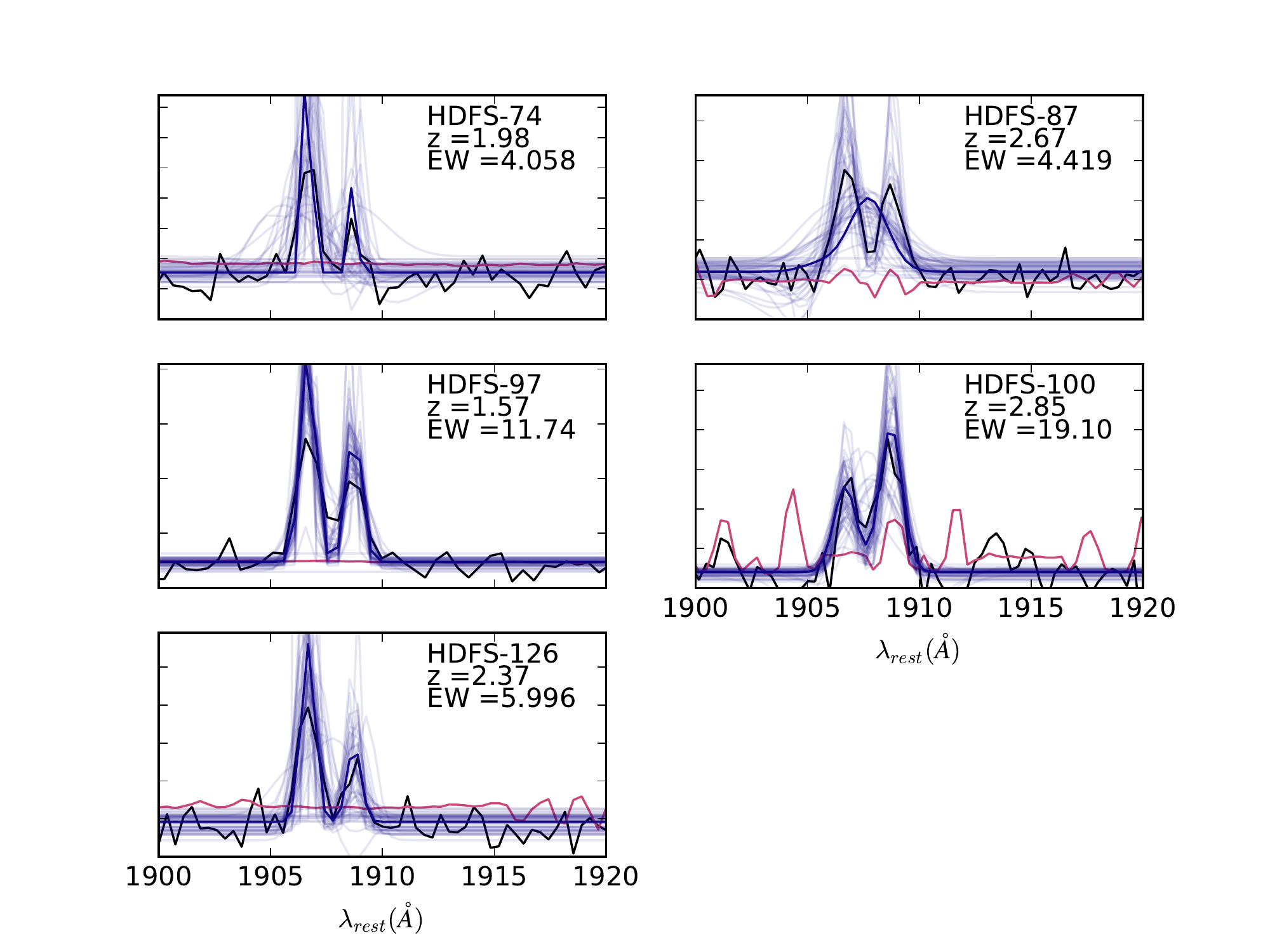}
\end{center}
\caption{MUSE spectra of the five C III]-emitters in the HDF-S.  Black denotes the measured flux and pink denotes the 1$\sigma$ error on the flux; the best-fit C III] doublet and 50 Monte Carlo iterations (performed on the spectrum with fluxes perturbed according to the measured errors) are shown with dark and light blue lines.}
\label{fig:hdfs}
\end{figure}

\begin{figure*}
\begin{center}
\includegraphics[width=.85\textwidth]{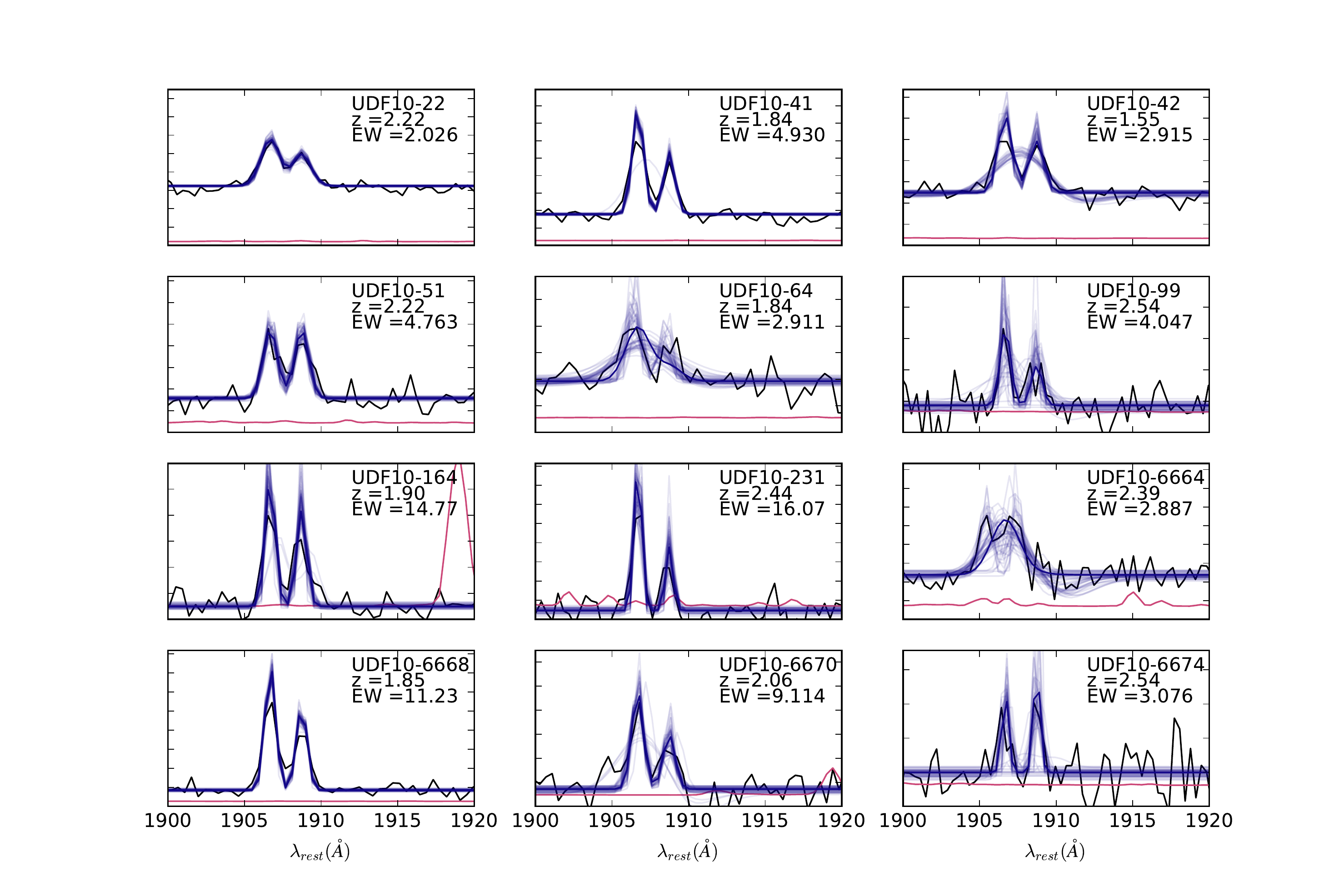}
\end{center}
\caption{MUSE spectra of the 12 C III]-emitters in the \udft.  Colors are identical to those in Figure \ref{fig:hdfs}.}
\label{fig:ciii}
\end{figure*}

Applying the criteria outlined in Section \ref{sec:lines}, we detect a total of 17 C III] emitters, 5 in the HDF-S (Figure \ref{fig:hdfs}) and 12 in the \udft (Figure \ref{fig:ciii}) summarized in Table \ref{tab:ciii}.  Even in the \udft where HST imaging shows a high spatial density of sources, all of our MUSE detections can be unambiguously attributed to a single source in both the \citet{2014ApJS..214...24S} and \citet{2015AJ....150...31R} catalogs.

If we were to relax our requirement on S/N to 1.5, we would have a sample of 29 emitters (20 in the \udft and 9 in the HDF-S).  We choose the $S/N >$ 3 threshold to ensure that the sample was clean of contaminants.  Since the flux uncertainties are larger in the red spectral regions because of the strong skyline contamination, lines in these regions are intrinsically less certain. Eight of the $1.5 < S/N < 3$ possible C III] emitters have redshifts $z > 2.9$, implying that C III] lies redward 7500 \AA.

\begin{table*}
\caption{MUSE C III] detections}              % title of Table
\label{tab:ciii}      % is used to refer this table in the text
\centering                                      % used for centering table
\begin{tabular}{lcccccccc}          
\hline\hline  
MUSE ID & RA & Dec & 3D-HST ID & R15 ID & z & C III] Flux & C III] EW & m$_{F606W}$ \\  & (degrees) & (degrees) & & & & ($10^{-20}$ erg s$^{-1}$ cm$^{-2}$) &  (\AA) & \\
\hline
  
  HDFS-74 & 338.22214 & -60.56792 & ... & ... & 1.984 & 179 $\pm$ 53.9 & 4.06 $\pm$ 1.75 & 25.48\\
HDFS-87 & 338.22859 & -60.56170 & ... & ... & 2.671 & 190 $\pm$ 56.3 & 4.42 $\pm$ 1.73 & 25.71\\
HDFS-97 & 338.23123 & -60.56595 & ... & ... & 1.571 & 702 $\pm$ 73.7 & 11.7 $\pm$ 2.34 & 25.91\\
HDFS-100 & 338.22970 & -60.55974 & ... & ... & 2.855 & 269 $\pm$ 40.8 & 19.1 $\pm$ 8.93 & 26.74\\
HDFS-126 & 338.24610 & -60.56492 & ... & ... & 2.372 & 179 $\pm$ 45.1 & 6.00 $\pm$ 2.50 & 26.03\\ \hline
UDF10-22 & 53.15447 & -27.77144 & 30443 & 8942 & 2.226 &  424 $\pm$ 17.9 & 2.03 $\pm$ 0.095 & 23.94\\
  UDF10-41 & 53.15288 & -27.77250 & 30427 & 9177 & 1.847 &  504 $\pm$ 17.2 & 4.93 $\pm$ 0.242 & 24.85\\
  UDF10-42 & 53.15829 & -27.77745 & 29293 & 9667 & 1.550 &  367 $\pm$ 20.7 & 2.92 $\pm$ 0.247 & 24.77\\
  UDF10-51 & 53.16518 & -27.78161 & 28248 & 10137 & 2.228 & 240 $\pm$ 16.5 & 4.76 $\pm$ 0.495 & 25.42\\
  UDF10-64 & 53.16001 & -27.77100 & 30782 & 9013 & 1.847 & 159 $\pm$ 27.6 & 2.91 $\pm$ 0.584 & 25.59\\
  UDF10-99 & 53.16308 & -27.78560 & 26992 & 6622 & 2.543 & 59.5 $\pm$ 11.6 & 4.05 $\pm$ 1.18 & 26.48\\
  UDF10-164 & 53.16935 & -27.78498 & 27151 & 6753 & 1.906 & 210 $\pm$ 16.0 & 14.8 $\pm$ 3.10 & 27.02\\
  UDF10-231 & 53.16210 & -27.77256 & 30259 & 9187 & 2.447 & 120 $\pm$ 14.7 & 16.1 $\pm$ 5.88 & 27.52\\
  UDF10-6664 & 53.16234 & -27.78444 & 27421 & 22123 & 2.394 & 116 $\pm$ 30.0 & 2.89 $\pm$ 0.631 & 25.68\\
  UDF10-6668 & 53.15342 & -27.78104 & 28278 & 7606 & 1.850 & 543 $\pm$ 16.2 & 11.2 $\pm$ 0.745 & 26.13\\
  UDF10-6670 & 53.16747 & -27.78183 & 28093 & 7257 & 2.069 & 191 $\pm$ 15.2 & 9.11 $\pm$ 1.96 & 26.54\\
  UDF10-6674 & 53.16656 & -27.77526 & 29650 & 9459 & 2.542 & 52.8 $\pm$ 10.1 & 3.08 $\pm$ 0.961 & 26.74\\

\end{tabular}
\tablefoot{Table of objects with detected C III] according to the criteria outlined in Section \ref{sec:lines}.  MUSE IDs come from \citet{2015AA...575A..75B} and Paper II for the HDFS and the \udft, respectively; 3D-HST IDs from the \citet{2014ApJS..214...24S} catalog refer specifically to the GOODS-S photometric catalog; ``R15'' refers to the \citet{2015AJ....150...31R} catalog.}
\end{table*}

In total, these 17 C III] detections constitute 3\% of the full sample of galaxies considered here.  The overall completeness of our search, and therefore the true fraction of C III] emitters, is difficult to establish.  As described in \citet{2015AA...575A..75B} and Paper II, the completeness of MUSE redshifts is mainly a function of continuum magnitude; the 50\% completeness of the \udft$~$(HDFS) is reached at $F775W \sim $ 26.5 ($F814W \sim$ 26).  Between $z\sim 1.5$ and $2.9$, C III] is the primary emission feature used in identifying redshifts, although brighter sources can have redshifts determined by absorption lines.  Above this redshift, many sources have redshifts that come from Ly-$\alpha$.  Unfortunately, it is nontrivial to determine the systemic redshift from Ly-$\alpha$ (cf. Verhamme et al. in prep) and hence it is difficult to ascertain the true flux distribution of C III] for objects that have strong Ly-$\alpha$, which would be used in the initial redshift determination, since we do not \textit{a priori} know exactly where C III] should be located.  In the most extreme cases, Ly-$\alpha$ can be offset by up to 1000 km s$^{-1}$ from the systemic redshift (Paper II).  These factors and the wavelength-dependent ability to observe C III] due to skyline contamination do not allow us to definitively obtain a corrected number density for C III] emitters in this data set.

Figures \ref{fig:hdfs} and \ref{fig:ciii} show the best-fit emission line model from \pf along with 50 of the Monte Carlo simulations for each spectrum.  It is clear that the Monte Carlo simulations are necessary not only for an accurate determination of the continuum level, as described in Section \ref{sec:lines}, but also for obtaining an accurate picture of the line profile.  Some objects that are clearly emission doublets, such as HDFS-87, have a best fit that is a single broad feature with no flux in a second component of the doublet.  In other cases, such as UDF10-42, \pf fits a single broad emission line as the blue component of C III] combined with a broad absorption line as the red component; \pf does allow emission line amplitudes to be negative.  This can be due to the initial conditions of the fitting or the nonlinear least squares algorithm settling at a local minimum for such a solution.  Since we do not impose restrictions on, for example, the amplitude ratio of the two components of C III], the Monte Carlo simulations provide some additional redundancy to ensure that all objects in our sample are definitively C III] emitters;  all emission lines shown here are, when incorporating the Monte Carlo simulations, well fit by an emission doublet with a line spacing corresponding to C III].

\subsection{Electron densities}
\label{sec:ciiidens}
\begin{figure}
\begin{center}
%~Dropbox/MUSE/CIII_tem_dens.py
\includegraphics[width=.45\textwidth]{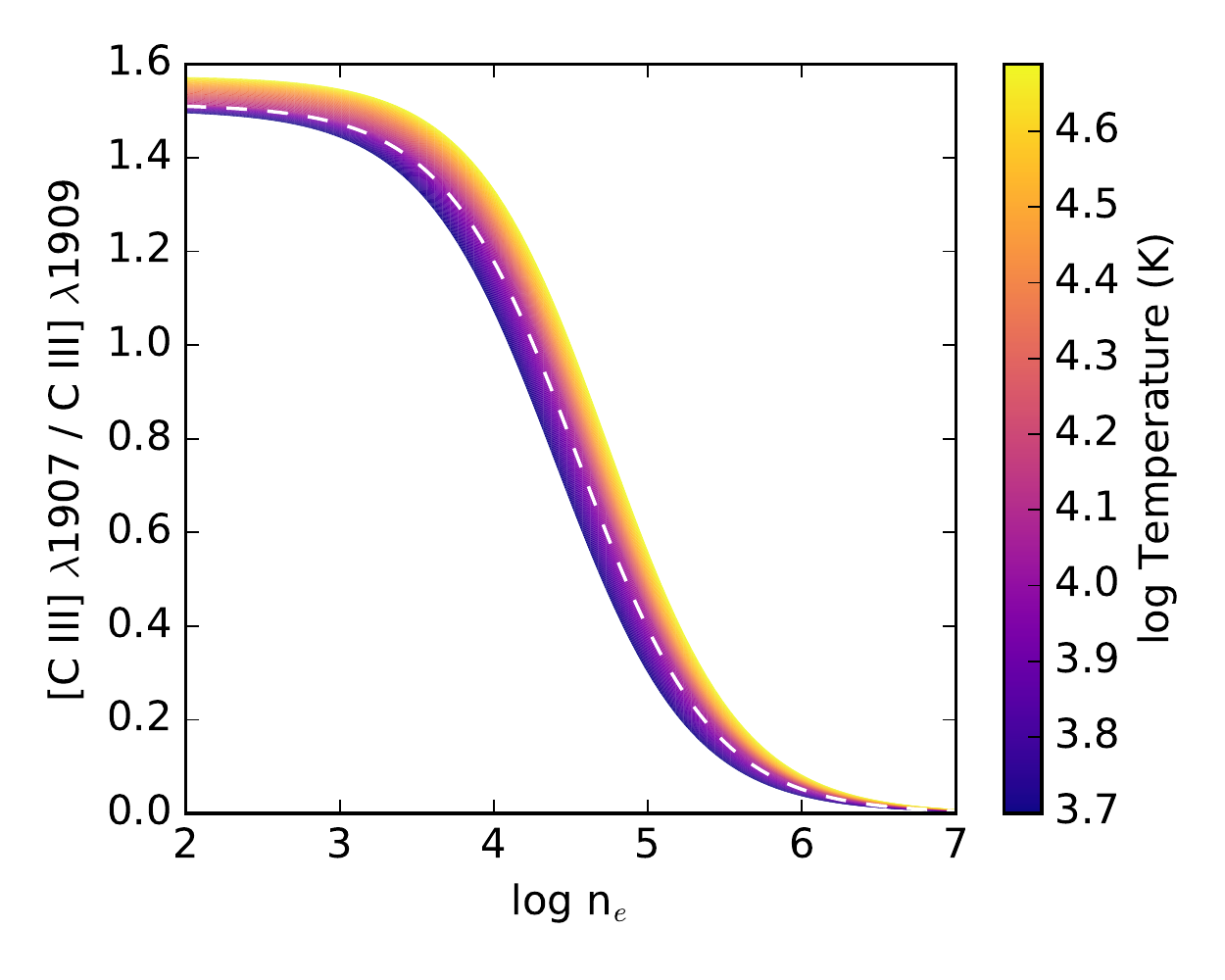}
\end{center}
\caption{Ratio of [C III] $\lambda$1907 to C III] $\lambda$1909 as a function of temperature and electron density, calculated via \texttt{PyNeb} \citep{2015AA...573A..42L}.  As noted in the text, the density derived from a fixed ratio is nearly independent of temperature; throughout we assume a temperature of 10,000 K (denoted by the dashed line).}
\label{fig:ciiitd}
\end{figure}

The ratio of [C III] 1907 \AA$ $ to C III] 1909 \AA$ $ can also be used as a tracer of the electron density in the interstellar medium, much like [O II] 3727/3729 \AA.  This is because the two lines come from the same ion at different energy levels with nearly the same excitation energy, hence the relative populations in each level are determined by the ratio of collision strengths \citep{2006agna.book.....O}.  This ratio as a function of temperature and electron density is shown in Figure \ref{fig:ciiitd}.  \citet{2016ApJ...816...23S} have found an elevated electron density in z$\sim$2.3 galaxies compared to local star-forming galaxies: using [O II] $\lambda \lambda$3727,3729 and [S II] $\lambda \lambda$6716,6731, they find mean electron densities of 225 cm$^{-3}$ and 290 cm$^{-3}$ compared to 26 cm$^{-3}$ locally.  While the C III] 1907/1909 ratio saturates to a ratio of 1.53 at densities below $\sim$10$^3$ cm$^{-3}$, we do observe electron densities in excess of this for at least some of our objects (four of the objects have values significantly greater than $10^3~$cm$^{-3}$ in Table \ref{tab:sed}), implying that the average densities in these C III] emitters could be much higher than locally.  

However, the ionization potentials required to create C III] and [O II] are very different: 24.4 and 13.6 eV, respectively.  If the HII regions that emit these photons are spherical with densities decreasing as $1/r^2$, then the measured electron densities from ratios of forbidden lines would be different as a function of radius.  In this model, lines with lower ionization potentials would measure the outer parts of the HII region due to luminosity weighting and hence [O II] would trace, on average, a much lower density interstellar medium than C III].  A more accurate comparison for C III]-derived densities would be [Cl III] $\lambda\lambda$5517,5537 (23.8 eV) or [Ar IV] $\lambda\lambda$4711,4740 (40.7 eV).  Using data from the Sloan Digital Sky Survey DR7 \citep{2000AJ....120.1579Y,2009ApJS..182..543A} and the \texttt{Platefit}-based pipeline from \citet{2008AA...485..657B}, we derive a sample of 165 galaxies with $>3$$\sigma$ detections in both [Cl III] lines and 280 galaxies with $>3$$\sigma$ detections in both [Ar IV] lines that are classified as star forming.  Using the \citet{1989ApJ...343..811S} conversions from the flux ratios to electron densities at $T = 10,000$ K, we found median (log) electron densities of log n$_e$=2.5$\pm$0.61 from [Cl III] and log n$_e$=3.4$\pm$0.46 from [Ar IV].  Considering the aforementioned model of HII regions, where the density estimate increases with increasing ionization potential, the increase from the local [O II] value of log n$_e$=1.4 to the [Cl III] and [Ar IV] values here is expected.  Our median value from C III] is log n$_e$=3.2, which is consistent with the $z\sim0$ [Ar IV] value determined here \cite[and still higher than the $z\sim2.3$ \lbrack O II\rbrack $ $ value of log n$_e$=2.4 from][]{2016ApJ...816...23S}.

Further investigations comparing the C III]-derived densities with, for example, [O II] derived densities for the same objects will shed more light onto this issue because cases where both doublets are observed in the same galaxy at high resolution are rare \citep{2010MNRAS.406.2616C, 2016MNRAS.456.4191P}.  This will require high-resolution near-IR spectroscopy in addition to MUSE, since there is no spectral overlap between [O II] and C III] with the wavelength coverage of MUSE.

\subsection{C III] and rest-frame optical emission lines}
\label{sec:grism}
\begin{figure*}
\begin{center}
\includegraphics[width=0.95\textwidth]{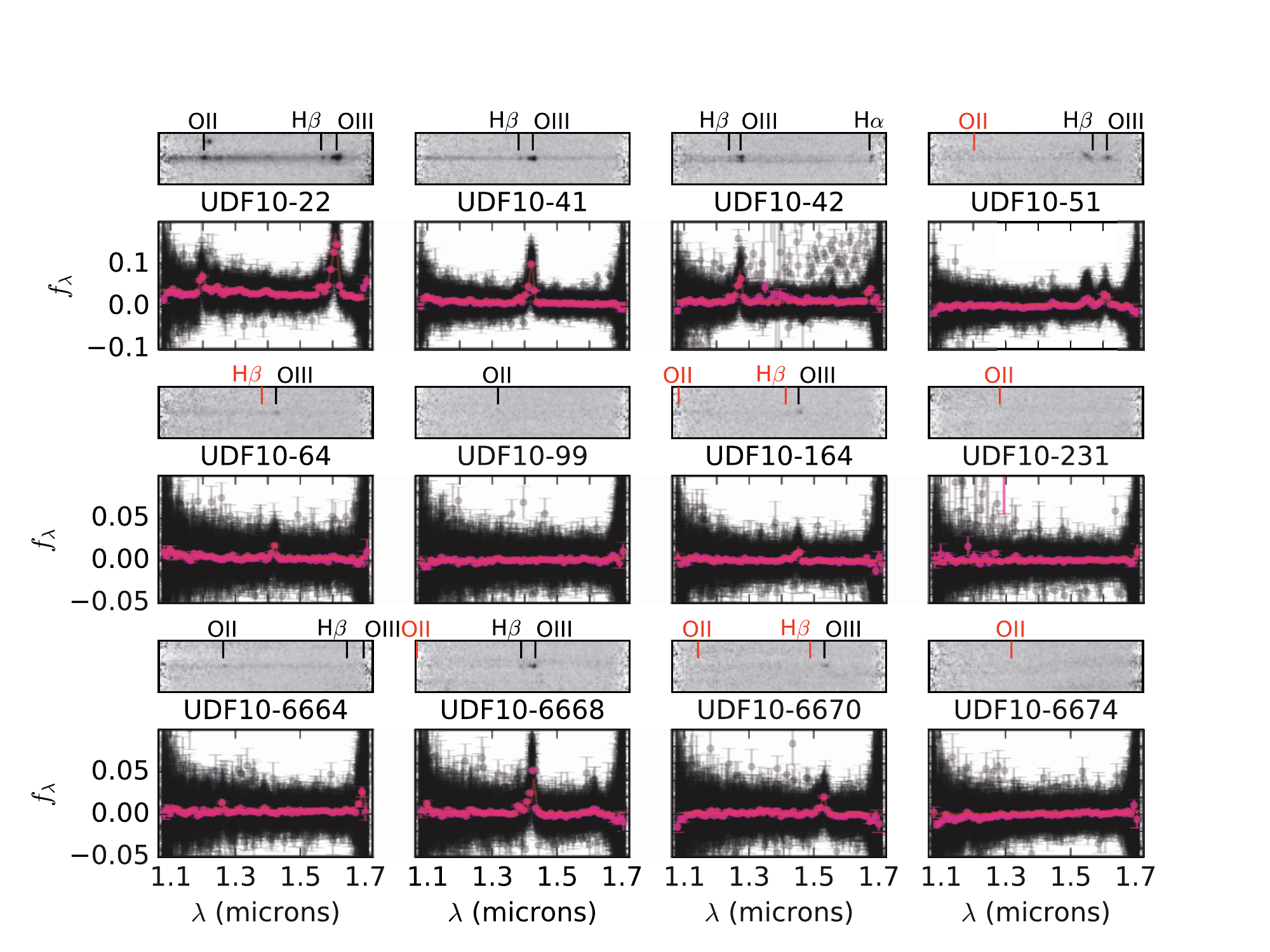}
%"compressed" using Preview->Export->Generic PDFX-3
\end{center}
\caption{HST WFC3/G141 grism spectra ($\sim$ 1.1$-$1.7 $\mu$m) for all C III] emitters in the MUSE \udft region.  For each object we show (top) the stacked two-dimensional grism spectrum and (bottom) the individual one-dimensional optimally extracted data points for each grism frame with the median and best-fit model shown in pink.  The vertical lines denote the positions of [O II], H$\beta$, [O III], and H$\alpha$ based on the MUSE redshift, with $>$ 3$\sigma$ detections in black and nondetections in red.  All objects in this area with a rest-frame [O III] EW in excess of 250 \AA, the so-called Extreme Emission Line Galaxies \citep{maseda14}, are C III] emitters.}
\label{fig:oiii}
\end{figure*}

Connecting the C III] emission with rest-frame optical emission lines, such as [O III] $\lambda$5007, is of particular interest to establish C III] as a useful redshift indicator in the epoch of reionization; numerous studies have shown that galaxies at $z\gtrsim6$ have strong nebular emission features \cite[e.g.,][]{2012ApJ...755..148G,2013ApJ...777L..19L,2014ApJ...784...58S, 2016ApJ...817...11H,2016ApJ...823..143R}.  These rest-frame optical features for the MUSE C III] emitters are redshifted into near-infrared wavelengths; in the redshift range $1.5 < z < 2.4$, we would be able to observe C III] with MUSE and [O III] with the G141 grism on HST/WFC3, which provides spectral coverage from 1.1 to 1.7 $\mu$m at $R\sim100$.  See Section \ref{sec:grismdat} for a description of the data.  Likewise, we have spectral coverage of C III] and [O II] at $1.8 < z < 3.6$\footnote{The spectral resolution of the G141 grism is too low to resolve the two components of the [O II] doublet, so we cannot obtain an estimate of the electron density from the 3727 to 3729 \AA $ $ ratio in these spectra.}.  Spectra for the MUSE C III] emitters are shown in Figure \ref{fig:oiii} and summarized in Table \ref{tab:grism}.

\begin{table*}
\caption{HST grism spectroscopic data for C III] emitters in the \udft}              % title of Table
\label{tab:grism}      % is used to refer this table in the text
\centering                                      % used for centering table
\begin{tabular}{lcccccccc}          
\hline\hline  
MUSE ID & HUDF ID &  z & [O III] EW & [O II] EW & H$\beta$ EW & H$\alpha$ EW & 12+log(O/H) & 12+log(O/H)\\  &  &  & (\AA) &  (\AA) & (\AA) & (\AA) & ($R_{23}$, lower) & ($R_{23}$, upper)\\
\hline
UDF10-22 & 2925 &  2.226 &  458 $\pm$ 9.52 & 88.4 $\pm$ 5.68 & 74.1 $\pm$ 4.10 & ... & 8.21 $\pm$ 0.052 & 8.44 $\pm$ 0.032\\
UDF10-41 & 2900 &  1.847 &  973 $\pm$ 38.2 & 86.0 $\pm$ 111 & 73.6 $\pm$ 10.3 & ... & 8.27 $\pm$ 0.223 & 8.42 $\pm$ 0.160 \\
UDF10-42 & 2455 &  1.550 &  444 $\pm$ 20.9 & ... & 62.1 $\pm$ 10.8 & 264 $\pm$ 23.6 & ... & ... \\
UDF10-51 & 2034  &  2.228 &  464 $\pm$ 44.4 & $<$ 279 & 145 $\pm$ 25.6 & ... & 7.42 $\pm$ 0.153 & 8.82 $\pm$ 0.051 \\
UDF10-64 & 3058 & 1.847 &  299 $\pm$ 32.3 & 444 $\pm$ 413 & 24.9 $\pm$ 20.9 & ... & ... & ... \\
UDF10-99 & 1622 & 2.543 &  ... & 120 $\pm$ 48.6 & ... & ... & ... & ... \\
UDF10-164 & 1698 & 1.906 &  1170 $\pm$ 292 & $<$ 3300 & 171 $\pm$ 74.3 & ... & 8.41 $\pm$ 0.398 & 8.31 $\pm$ 0.280 \\
UDF10-231 & 2960 & 2.447 &  ... & $<$ 641 & ... & ... & ... & ... \\
UDF10-6664 & 1727 & 2.394 &  473 $\pm$ 119 & 88.8 $\pm$ 14.2 & 65.2 $\pm$ 20.4 & ... & 8.44 $\pm$ 0.325 & 8.28 $\pm$ 0.227\\
UDF10-6668 & 2090 & 1.850 &  937 $\pm$ 87.6 & $<$ 113 & 180 $\pm$ 30.9 & ... & ... & ...\\
UDF10-6670 & 2018 & 2.069 &  535 $\pm$ 66.7 & 46.1 $\pm$ 34.1 & 39.9 $\pm$ 27.6 & ... & 8.87 $\pm$ 0.857 & 8.09 $\pm$ 0.544\\
UDF10-6674 & 2720 & 2.542 &  ... & $<$ 209 & ... & ... & ... & ...\\

\end{tabular}
\tablefoot{Grism spectroscopic data for the C III] emitters in the UDF10.  HUDF IDs come from the \citet{2013arXiv1305.2140V} catalog.  ``[O II]'' refers to the combined [O II] $\lambda\lambda$3727,3729 doublet and ``[O III]'' refers to the combined [O III] $\lambda\lambda$4959,5007 doublet.  The redshift coverage of C III] in MUSE and [O III] in the G141 grism is $1.4 \lesssim z \lesssim 2.4$.  All EWs are quoted in the rest frame.  Upper limits for [O II] are based on the broadband $J_{F125W}$ continuum level and the brighter of the 3$\sigma$ flux measurement from the spectrum or the flux limit of the stacked grism spectra \cite[3-$\sigma~\sim$ 3.9 $\times$ 10$^{-18}$ erg s$^{-1}$ cm$^{-2}$;][]{2013ApJ...765L...2B}.  Values of $12 + log (O/H)$ (gas-phase metallicity) are estimated using the $R_{23}$ ratio [([O II] + [O III])/H$\beta$] and the calibration of \citet{1999ApJ...514..544K} for both the upper and lower branches of the $R_{23}$ parameter.}
\end{table*}

With these slitless/integral field unit (IFU) data, we are in the position to study the relationship between C III] and [O III] emission without the need for standard preselections (e.g., photometric redshift) that are necessary for targeted spectroscopic studies.  While previously the strengths of rest-frame optical lines in C III] emitters could only be estimated via excesses in broadband photometry \cite[e.g.,][]{2014MNRAS.445.3200S,2017arXiv170104416A}, here we show that all MUSE C III] detections at $1.5 < z < 2.4$ have significant detections of [O III] and most at $z > 1.9$ have detections of [O II].

\begin{figure}
\begin{center}
\includegraphics[width=0.45\textwidth]{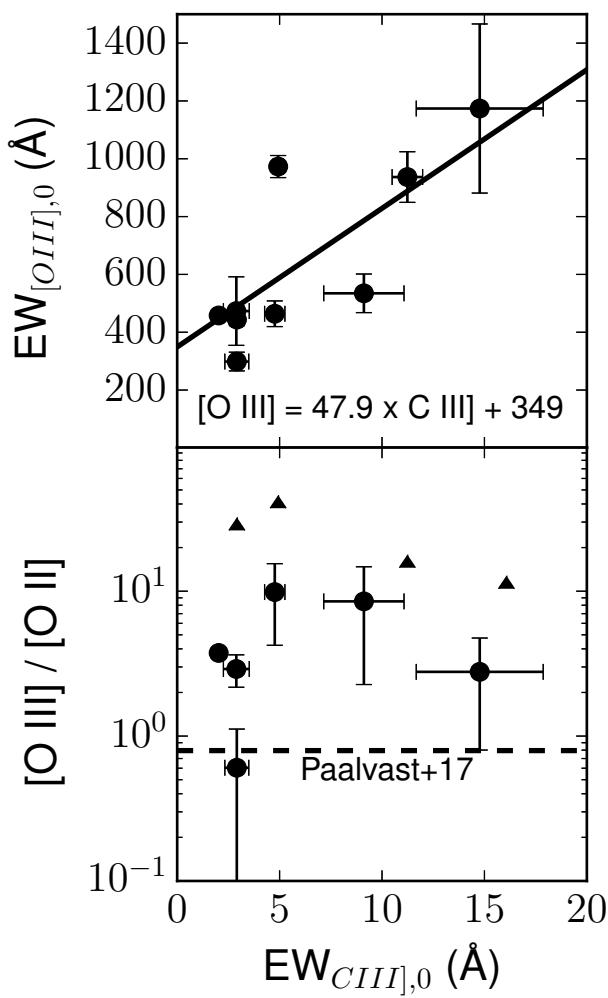}
%muse2: ITERA/ciii_oii.py
\end{center}
\caption{Rest-frame C III] $\lambda \lambda$1907,1909 EW from MUSE spectroscopy vs. (top) rest-frame [O III] $\lambda\lambda$4959,5007 EW and (bottom) the ratio of [O III] to [O II] flux, all from WFC3/G141 grism spectroscopy.  Lower limits in the bottom panel are based on the grism flux limit \cite[3$\sigma~\sim$ 3.9 $\times$ 10$^{-18}$ erg s$^{-1}$ cm$^{-2}$;][]{2013ApJ...765L...2B} when [O II] is not detected spectroscopically.  A (linear) relation is expected between the EWs of C III] and [O III] since the intense ionizing radiation fields necessary for exciting C III] are also expected to generate the collisional excitations needed for [O III] emission; here we find a linear relation between the rest-frame EWs of $[O III] = 47.9 \times C III] + 349$.  The [O III] to [O II] ratio traces two different ionization levels of the oxygen gas and therefore serves as a measure of the intensity of the radiation field within the galaxy.  The dashed line in the bottom panel shows the median ratio of 0.794 from Paalvast et al. (in prep.) for MUSE star-forming galaxies at $0.28 < z < 0.85$.  The C III] emitters predominantly have higher [O III] to [O II] ratios than the lower-z sample.}
\label{fig:oiiiciii}
\end{figure}

In the top panel of Figure \ref{fig:oiiiciii} we show the relationship between the rest-frame EWs of C III] and [O III].  Remarkably, the relationship between the two rest-frame EWs can be approximated by a linear function, $[O~III] = 47.9 \times C~III] + 349$.  \citet{2014ApJ...784...58S} have estimated that the average rest-frame EW of [O III] + H$\beta$ at z $\sim$ 6 is 637 \AA; if H$\beta$ is 1:8 of the total combined EW \citep{2011ApJ...742..111V} then the EW$_{\lambda\lambda 4959,5007}$ is 557 \AA, implying a C III] EW of 4.3 \AA.  

Broadband photometry for C III] emitters in the \citet{2014MNRAS.445.3200S} and \citet{2017arXiv170104416A} samples also show plausible signs of contamination from rest-frame optical emission lines such as [O III].  Such a relation is expected in photoionization models since the high nebular temperatures, ionization parameters, and ionizing radiation from a young stellar population required to generate large C III] fluxes also generate large fluxes in collisionally excited forbidden lines such as [O III] \cite[e.g.,][]{2016ApJ...833..136J}; the ionization energy for C III] is 24.4 eV and for [O III] is 35.1 eV.  The converse that strong [O III] flux is associated with strong C III] flux is not necessarily true since the excitation potentials of the lines are different, i.e., $\sim$7 eV for C III] and $\sim$1 eV for [O III].  We can invert the problem and study the C III] properties of [O III] emitters selected from the grism data.  As shown in \citet{Maseda13,maseda14}, galaxies selected on the basis of high-EW [O III] (and H$\alpha$) at these redshifts are nearly always low-mass, low-metallicity, bursty star-forming galaxies.  All objects in the \udft footprint that have an ``extreme'' [O III] equivalent width (i.e., $>$ 250 \AA) have detections of C III] in MUSE.  This is in broad agreement with the results shown in Section \ref{sec:sed}, where the C III] emitters have higher sSFRs than nonemitters at the same redshifts.

\begin{figure}
\begin{center}
\includegraphics[width=0.48\textwidth]{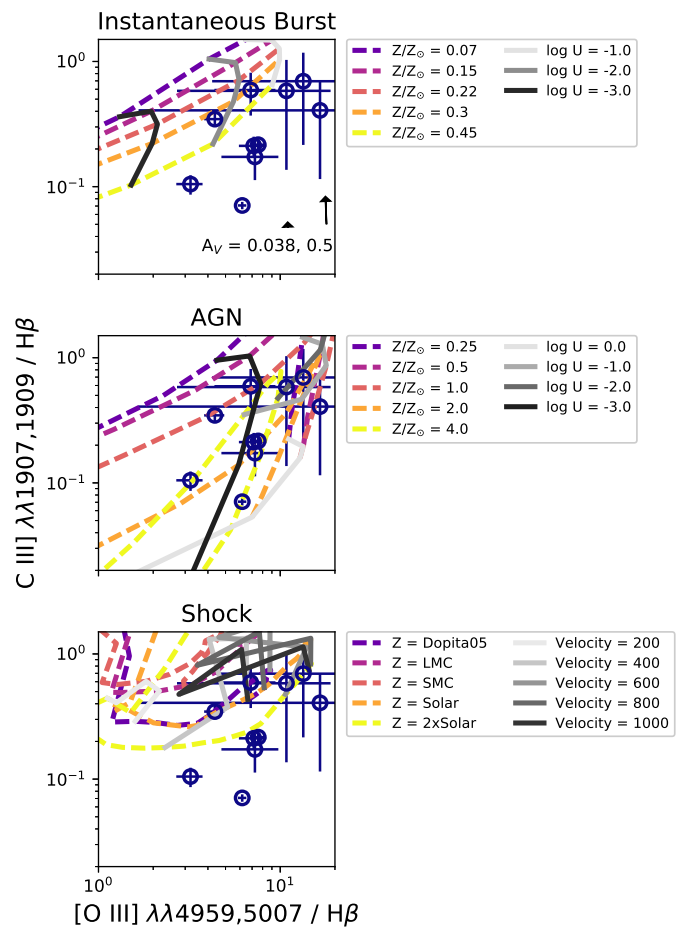}
\end{center}
\caption{Comparison of the expected ratios of [O III] and C III] to H$\beta$ fluxes from models of star formation, AGN, and shock excitation, color coded by metallicity, compared to our measurements from the combination of MUSE and G141 spectroscopy.  The star-forming grid shows the fiducial model from \citet{2016ApJ...833..136J}, the AGN grid shows the dust-free isochoric narrow line region models from \citet{2004ApJS..153...75G}, and the shock grid indicates the fully radiative shock plus precursor model from \citet{2008ApJS..178...20A}: see the text for more details. The units of (shock) velocity indicate km s$^{-1}$ and $U$ indicates the dimensionless ionization parameter, volume averaged for the SF grids and the value at the inner edge of the gaseous nebula for the AGN grids.  All metallicity values are expressed as a fraction of the solar metallicity.  The two vectors show the effect of dust reddening with an $A_V$ of 0.038 (the median of our C III] emitters) and 0.5 magnitudes.  While dust attenuation is a contributing factor, in general star formation models do not reproduce the observed distribution of line ratios, namely the high [O III]/H$\beta$ ratios in objects with low C III]/H$\beta$ ratios; such a large offset could imply that either more extreme photoionization models, different nebular parameters (such as a lower C/O ratio), or nonstellar forms of excitation are required.}

\label{fig:diagnostic}
\end{figure}

In Figure \ref{fig:diagnostic} we show the relationship between the fluxes (normalized by the H$\beta$ flux).  The star-forming grid shows the fiducial model from \citet{2016ApJ...833..136J} for an instantaneous burst with an age of 1 Myr, $n_H$ = 100 cm$^{-3}$, a C/O ratio of 0.2, and the BPASS spectral synthesis models \citep{2016MNRAS.456..485S}; the AGN grid shows the dust-free isochoric narrow line region models from \citet{2004ApJS..153...75G} with a power-law index $\alpha$ = -1.4 and $n_H$ = 1000 cm$^{-3}$; the shock grid indicates the fully radiative shock plus precursor model from \citet{2008ApJS..178...20A} with a magnetic parameter $B/n^{1/2}$ of 1 $\mu$G cm$^{3/2}$ for five different atomic abundance sets (including the set from \citeauthor{2005ApJ...619..755D} \citeyear{2005ApJ...619..755D}; see Table 1 of \citeauthor{2008ApJS..178...20A} \citeyear{2008ApJS..178...20A} for details) and a preshock density of 1 cm$^{-3}$.  The AGN and shock model grids were created using the \texttt{ITERA} \citep{2010NewA...15..614G} code.  

In general, the photoionization models are not well constrained by observations owing to the small existing sample sizes.  While a full treatment with a larger sample of C III] emitters will be presented by Maseda et al. (in prep.), we show the [O III]/H$\beta$ versus C III]/H$\beta$ diagnostic in Figure \ref{fig:diagnostic}.  In the \citet{2016ApJ...833..136J} tracks, which include the effects of binary stars and stellar rotation via the BPASS \citep{2016MNRAS.456..485S} spectral synthesis models, the high [O III]/H$\beta$ ratios can only be produced by very hard (log $U~\sim$ -1) ionizing spectra.  The ratio of [O III] to [O II] flux to first order constrains the ionization state of the gas, although this ratio also depends somewhat on metallicity.  In the bottom panel of Figure \ref{fig:oiiiciii}, we see that the highest [O III] to [O II] ratios are all poorly constrained, but on average the C III] emitters have higher [O III] to [O II] ratios than the vast majority of star-forming galaxies at $0.28 < z < 0.85$ (Paalvast et al. in prep.).  Elevated [O III] to [O II] ratios indicate that C III] emitters have more intense ionizing radiation fields, since the ratio traces two different ionization levels for the same atom.

In general, the star formation models lie in a different region of parameter space than our observations here; our C III] emitters have high [O III]/H$\beta$ ratios and low C III]/H$\beta$ ratios.  This is not entirely due to dust extinction, which would move points to the upper left, since we show in the following Section that these objects have very low values of $A_V$, $\sim$0.038 magnitudes.\ The $A_V$ values of $\sim$0.5 in a \citet{2000ApJ...533..682C} extinction law would be required to change the observed C III] to H$\beta$ ratio to lie on the star formation grids; these are much larger than the $A_V$ values observed here.  Even though the model grids are calculated at a density of 100 cm$^{-3}$ and we observe higher densities in Section \ref{sec:ciiidens}, \citet{2016ApJ...833..136J} point out that increasing the density over several orders of magnitude only slightly enhances C III].  The discrepancy in line ratios could imply that either more extreme photoionization models, different nebular parameters (such as a lower C/O ratio), or nonstellar forms of excitation are required.

While the high [O III]/H$\beta$ could be evidence for some contribution by  AGN or shocks, we stress that such models have many free parameters such as C/O abundance and density, so additional constraints from other UV emission lines, such as C IV, He II, [Si III], and [O III] $\lambda\lambda$1661,1666 will be critical to disentangle the different scenarios \citep{2016MNRAS.456.3354F}.  As in \citet{2017arXiv170104416A}, none of our C III] emitters in the \udft have detections in the deep 4 Ms \textit{Chandra} X-ray catalogs \citep{2011ApJS..195...10X} and are plausibly not luminous unobscured AGN.  In any case, large [O III]/H$\beta$ ratios (in excess of 7:1) are common in strongly star-forming, low-metallicity galaxies at $z\sim2$ \cite[e.g.,][]{2011ApJ...742..111V,2016ApJ...832..171T}.

UDF10-22 has the lowest C III]/H$\beta$ ratio in Figure \ref{fig:diagnostic} which, combined with its high [O III]/H$\beta$ ratio, is not consistent with the star formation grids even when taking into account its $A_V$ value of 0.5 magnitudes. It also has the highest measured UV luminosity, brightest $F606W$ magnitude, largest effective radius, highest stellar mass, highest star formation rate, and reddest $\beta$ slope of all C III] emitters in this sample.  This object has signatures of AGN activity in an archival 1h VLT/X-Shooter spectrum (093.A-0882(A); PI: Atek) via asymmetric [O III] emission lines and an [O III]/H$\beta$ versus [N II]/H$\alpha$ value that is consistent with an AGN; UV emission line diagnostics featuring C III], C IV $\lambda\lambda$1548,1550, He II $\lambda$1640, [Si III] $\lambda\lambda$ 1883, 1892, and O III] $\lambda\lambda$1661,1666 \citep{2016MNRAS.456.3354F} are consistent with a ``composite'' object with some AGN contribution (A. Plat et al. in prep).  The object is also detected in deep 1.2 mm-continuum observations from ALMA \cite[XDFU-2370746171;][]{2016arXiv160605280B}.

%\subsubsection{Gas-phase metallicties}
%\label{sec:metal}
In Table \ref{tab:grism} we include estimates for the gas-phase oxygen abundance (a proxy for total metallicity) using the calibration of \citet{1999ApJ...514..544K} based on the ``$R_{23}$'' ratio, ([O II] + [O III]) / H$\beta$.  The mapping from this ratio to the gas-phase metallicity is double valued, meaning that any one value of $R_{23}$ can pertain to a low or high metallicity.  These two solutions are referred to as the ``lower'' and ``upper'' branches.  The fact that a majority (4/7) of our values are consistent within 1$\sigma$ between the two branches comes from the high $R_{23}$ values observed here, often in excess of 10.  High $R_{23}$ values lie in the crossover region between the upper and lower branches and result in metallicity estimates that are similar or inverted for the two branches.  More precise gas-phase oxygen abundances would be valuable and can potentially be obtained by combining these line fluxes with, for example, constraints on the electron temperature using O III] $\lambda$1666 and [O III] $\lambda$5007.  Such an analysis will be the topic of future work.

\begin{figure}
\begin{center}
%schenk: MUSE/ciii/metallicity_singlepanel.py
\includegraphics[width=0.45\textwidth]{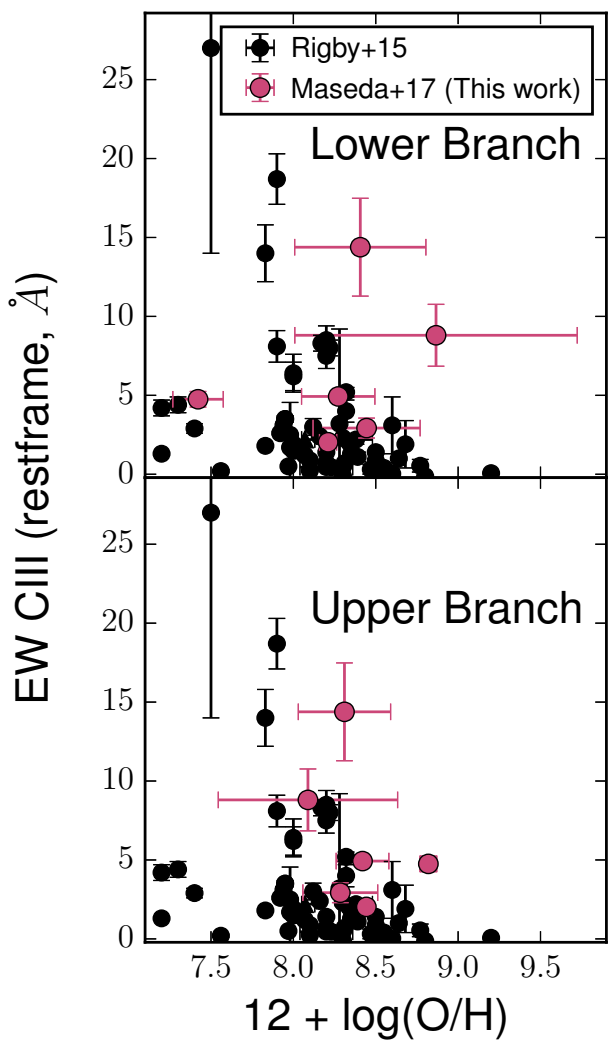}
\end{center}
\caption{Rest-frame C III] EW vs. gas-phase metallicity.  Black points represent the $z\sim0$ and $z\sim2$ literature compilation of \citet{2015ApJ...814L...6R} with triangles denoting lower limits on EW, and pink points represent the MUSE galaxies.  Based on this data, \citet{2015ApJ...814L...6R} claim that metallicity sets an envelope to the C III] equivalent width, i.e., low metallicity is a necessary but not sufficient condition for high C III] equivalent width.  Regardless of whether we adopt the upper or lower $R_{23}$ branch metallicities, we cannot rule out such a claim with the MUSE objects.}
\label{fig:oh}
\end{figure}

In Figure \ref{fig:oh} we show our metallicity estimates versus the measured EW of C III].  Also shown are values from the literature compilation of \citet{2015ApJ...814L...6R} at $z\sim0$ and $z\sim2$.  \citet{2015ApJ...814L...6R} suggest that metallicity sets an envelope on the EW of C III] and EWs in excess of $\sim$5 \AA$~$are only present at low metallicities, below $12 + log (O/H) \sim 8.4$ or $\sim 0.5~Z_{\odot}$.  \citet{2010ApJ...719.1168E} use photoionization models to conclude that C III] peaks in intensity at a metallicity of $\sim 0.2~Z_{\odot}$ and decreases at both higher and lower metallicities.  In effect this implies that low metallicity is a necessary but not sufficient condition for a high C III] EW.  While the lack of other spectroscopic information, such as [N II]/[O II], prevents us from determining if individual galaxies lie on the upper or lower $R_{23}$ branch, our highest EW C III] emitters are consistent with this assertion regardless of the branch.

\subsection{Derived parameters and comparisons to the full sample}
\label{sec:sed}

\begin{figure*}
\begin{center}
\includegraphics[width=.9\textwidth]{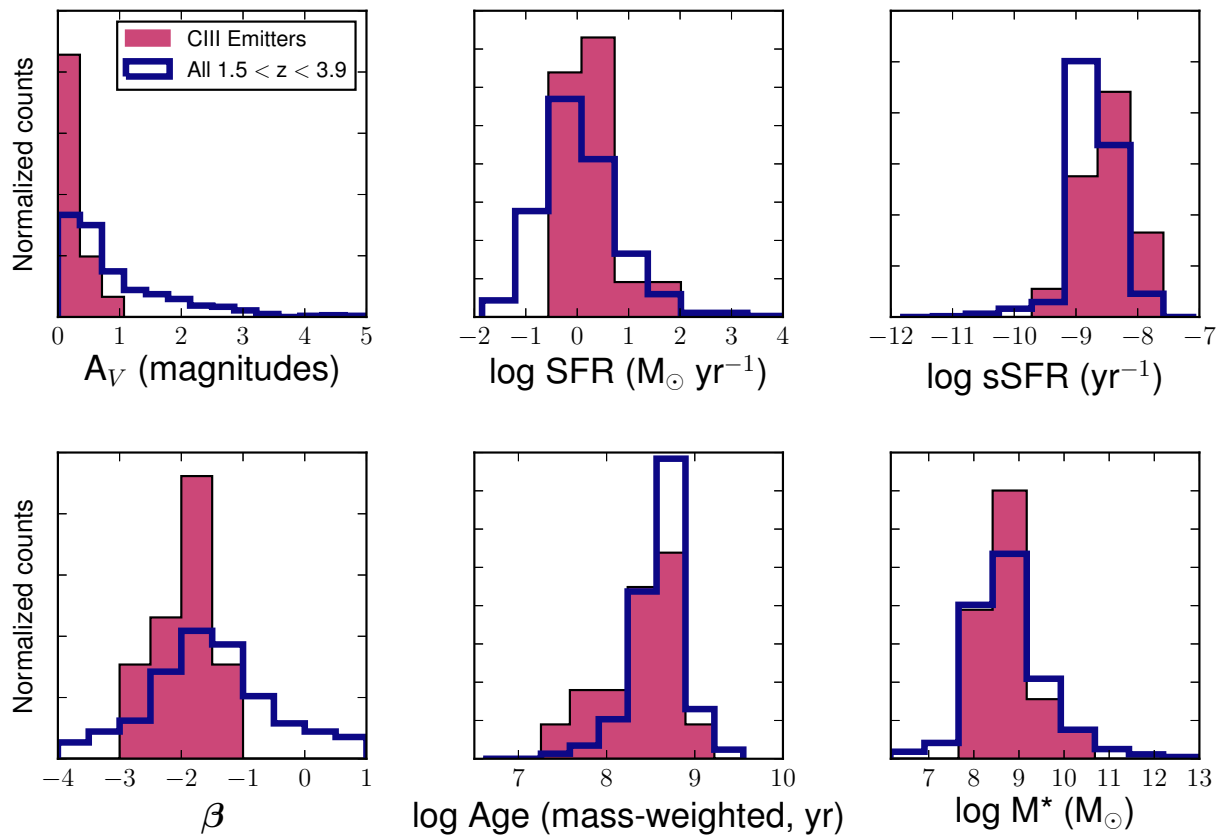}
\end{center}
\caption{Normalized histograms of output quantities from \texttt{MAGPHYS}, all derived using the HDF-S and 3D-HST photometry.  The values for C III] emitters shown in pink and for the full photometric sample in the redshift range shown in blue. Each value shown is the median of the probability distributions produced by \texttt{MAGPHYS} when marginalizing over all other parameters.  Star formation rates and specific star formation rates are averaged over the past 0.1 Gyr.}
\label{fig:exp}
\end{figure*}

Given the blind and untargeted nature of our IFU spectroscopy, we can compare the values of the SED-derived parameters for our C III] emitters and all other galaxies at 1.49 $< z <$ 3.90, performed with \magphys as described in Section \ref{sec:sedfit}, in the same redshift range to look for global differences.  Normalized histograms from \magphys for the C III] and total samples are shown in Figure \ref{fig:exp}.  The values are the median of the output probability distributions from \texttt{MAGPHYS} when marginalizing over all other parameters.  Since the stacks are normalized, the shape of the distributions reveals differences between the two samples.

\begin{table*}
\caption{Physical parameters for MUSE C III] emitters in the HDFS and \udft}              % title of Table
\label{tab:sed}      % is used to refer this table in the text
\centering  
\begin{tabular}{lccccccc}
\hline \hline
  \multicolumn{1}{c}{ID} &
  \multicolumn{1}{c}{log L$_{UV}$} &
  \multicolumn{1}{c}{log SFR} &
  \multicolumn{1}{c}{log M$_{\star}$} &
  \multicolumn{1}{c}{log Age} &
  \multicolumn{1}{c}{Av} &
  \multicolumn{1}{c}{$\beta$} &
  \multicolumn{1}{c}{log $n_e$}\\
  \multicolumn{1}{c}{} &
  \multicolumn{1}{c}{(L$_{\odot}$)} &
  \multicolumn{1}{c}{(\msol~yr$^{-1}$)} &
  \multicolumn{1}{c}{(\msol)} &
  \multicolumn{1}{c}{(yr)} &
  \multicolumn{1}{c}{(mag)} &
  \multicolumn{1}{c}{} &
  \multicolumn{1}{c}{(cm$^{-3}$)}\\
\hline
  HDFS-74 & 10.0 & 0.072$^{+0.000}$$_{-0.000}$ &  8.07$^{+0.000}$$_{-0.000}$ & 7.79$^{+0.000}$$_{-0.000}$ & 0.088$^{+0.000}$$_{-0.000}$ & ... & ... \\
  HDFS-87 & 10.2 & 0.212$^{+0.000}$$_{-0.000}$ & 8.21$^{+0.000}$$_{-0.000}$ & 7.79$^{+0.000}$$_{-0.000}$ & 0.088$^{+0.000}$$_{-0.000}$ & ... & ... \\
  HDFS-97 & 9.71 & -0.068$^{+0.045}$$_{-0.035}$ & 8.73$^{+0.155}$$_{-0.290}$ & 8.69$^{+0.195}$$_{-0.230}$ & 0.013$^{+0.025}$$_{-0.000}$ & ... & 3.0 $\pm$ 0.90 \\
  HDFS-100 & 9.41 & 0.422$^{+0.390}$$_{-0.505}$ & 9.71$^{+0.205}$$_{-0.220}$ & 9.08$^{+0.205}$$_{-0.360}$ & 1.04$^{+0.575}$$_{-0.625}$ & ... & 4.4 $\pm$ 0.60\\
  HDFS-126 & 9.92 & 0.152$^{+0.000}$$_{-0.080}$ & 8.26$^{+0.000}$$_{-0.000}$ & 7.92$^{+0.000}$$_{-0.000}$ & 0.038$^{+0.000}$$_{-0.000}$ & -1.99 $\pm$ 0.004 & ... \\
    UDF10-22 & 11.2 & 1.43$^{+0.000}$$_{-0.000}$ & 10.1$^{+0.000}$$_{-0.010}$ & 8.43$^{+0.000}$$_{-0.000}$ & 0.463$^{+0.000}$$_{-0.000}$ & -1.18 $\pm$ 0.035 & 2.9 $\pm$ 0.47\\
    UDF10-41 & 10.4 & 0.637$^{+0.020}$$_{-0.045}$ & 9.08$^{+0.050}$$_{-0.050}$ & 8.36$^{+0.060}$$_{-0.090}$ & 0.138$^{+0.050}$$_{-0.025}$ & -1.61 $\pm$ 0.038 & < 3 \\
      UDF10-42 & 10.2 & 0.427$^{+0.370}$$_{-0.000}$ & 9.06$^{+0.000}$$_{-0.140}$ & 8.60$^{+0.000}$$_{-0.610}$ & 0.038$^{+0.375}$$_{-0.000}$ & -1.97 $\pm$ 0.038 & 3.5 $\pm$ 0.31\\
  UDF10-51 & 10.2 & 0.397$^{+0.000}$$_{-0.000}$ & 9.30$^{+0.000}$$_{-0.000}$ & 8.80$^{+0.000}$$_{-0.000}$ & 0.038$^{+0.000}$$_{-0.000}$ & -1.91 $\pm$ 0.110 & 3.9 $\pm$ 0.20\\
    UDF10-64 & 10.0 & 0.202$^{+0.035}$$_{-0.010}$ & 8.67$^{+0.035}$$_{-0.000}$ & 8.41$^{+0.000}$$_{-0.020}$ & 0.038$^{+0.025}$$_{-0.000}$ & -2.34 $\pm$ 0.058 & ... \\
  UDF10-99 & 10.0 & 0.177$^{+0.000}$$_{-0.000}$ & 8.09$^{+0.000}$$_{-0.000}$ & 7.55$^{+0.000}$$_{-0.000}$ & 0.363$^{+0.000}$$_{-0.000}$ & -1.69 $\pm$ 0.421 & ... \\
     UDF10-164 & 9.36 & -0.413$^{+0.015}$$_{-0.000}$ & 8.42$^{+0.000}$$_{-0.315}$ & 8.69$^{+0.000}$$_{-0.225}$ & 0.013$^{+0.000}$$_{-0.000}$ & -2.93 $\pm$ 0.474 & 3.4 $\pm$ 0.56\\
     UDF10-231 & 9.34 & -0.388$^{+0.070}$$_{-0.055}$ & 8.07$^{+0.195}$$_{-0.250}$ & 8.46$^{+0.145}$$_{-0.535}$ & 0.038$^{+0.075}$$_{-0.025}$ & -2.50 $\pm$ 0.532 & ...\\
     UDF10-6664 & 10.7 & 0.882$^{+0.000}$$_{-0.005}$ & 9.02$^{+0.000}$$_{-0.000}$ & 8.01$^{+0.000}$$_{-0.000}$ & 0.713$^{+0.000}$$_{-0.000}$ & -1.84 $\pm$ 0.091 & ... \\
        UDF10-6668 & 9.73 & -0.043$^{+0.110}$$_{-0.000}$ & 8.79$^{+0.000}$$_{-0.090}$ & 8.69$^{+0.000}$$_{-0.085}$ & 0.013$^{+0.025}$$_{-0.000}$ & -2.23 $\pm$ 0.091 & 2.9 $\pm$ 0.32\\
       UDF10-6670 & 9.75 & -0.023$^{+0.000}$$_{-0.080}$ & 8.89$^{+0.000}$$_{-0.160}$ & 8.88$^{+0.000}$$_{-0.195}$ & 0.113$^{+0.000}$$_{-0.100}$ & -2.19 $\pm$ 0.316 & < 3 \\               
  UDF10-6674 & 10.2 & 0.047$^{+0.255}$$_{-0.055}$ & 8.52$^{+0.150}$$_{-0.055}$ & 8.41$^{+0.195}$$_{-0.325}$ & 0.038$^{+0.175}$$_{-0.025}$ & -1.22 $\pm$ 0.360 & 4.1 $\pm$ 0.78\\

\end{tabular}
\tablefoot{With the exception of $n_e$, all parameters are derived from broadband SED fits using \magphys \citep{2008MNRAS.388.1595D}.  Ultraviolet luminosities are the unattenuated values at rest-frame 1900 \AA, as measured from the best-fit \magphys SED.  Values and quoted uncertainties for \magphys parameters (SFR, $M_{\star}$, $Age$, and $A_V$) denote the median and shortest 68\% confidence interval centered on the median.  Star formation rates are averaged over the past 0.1 Gyr and ages are mass weighted. The value  $n_e$ is the electron density measured from the ratio of C III] 1907 to 1909 \AA $ $ \citep{2006agna.book.....O} at $T = $10,000 K when the signal to noise in each of the individual components is $> 3$$\sigma$; a measured 1907/1909 ratio in excess of the value in the low-density limit of 1.53 implies that the actual electron density is < 10$^3$ cm $^{-3}$.}
\end{table*}

In order to quantify the differences in the distributions shown in Figure \ref{fig:exp}, we use a two-sided Kolmogorov-Smirnov test.  By applying this test to the distributions of the median values, we can reject the null hypothesis that the C III] and total distributions are drawn from the same underlying distribution to better than 96\% for SFR, Age, $A_V$, and $\beta$, and to better than 92.5\% for the median values of sSFR.  Only the distributions in stellar mass cannot be conclusively differentiated.  Specifically, the \textit{p} values are 0.026 for SFR, 0.039 for Age, 4.0 $\times$ 10$^{-6}$ for $A_V$, 0.039 for $\beta$, 0.073 for sSFR, and 0.77 for stellar mass.

Figure \ref{fig:exp} shows that C III] emitters tend to have lower $A_V$ values than the total population and never have extinction values in excess of $\sim$ 1 magnitudes.  A similar picture emerges when looking at the rest-frame UV continuum slope $\beta$, where the C III] emitters are distributed around $\beta \sim -2$ and do not extend to positive (red) values.  This relation between C III] and the UV continuum is also found in \citet{2016arXiv161206866D}, who find stronger C III] in galaxies that are bluer in $(U~-~B)$.  C III] emitters have lower ages and higher star formation rates, leading to a higher average specific star formation rate.

Since we are capable of detecting C III] out to $z\sim3.9$, the histograms in Figure \ref{fig:exp} include objects out to $z\sim3.9$.  Nearly 30\% of the total sample (195 galaxies) is at a redshift higher than our highest redshift C III] emitter, z$\sim$2.9.  If galaxies at higher redshifts are younger, bluer, and more vigorously star forming than galaxies at lower redshifts, then the differences mentioned between the distributions at a fixed redshift could be even stronger.  Ultimately, larger samples of C III] emitters will be required to definitively quantify the average offset in these quantities.

As pointed out in, for example, \citet{2009AA...502..423S}, contamination of broadband photometry by high-EW emission lines can lead to systematic errors in determining galaxy properties through SED fitting.  In Section \ref{sec:grism}, we demonstrate that all C III] emitters in the \udft are high-EW [O III] and/or [O II] emitters and similarly that all of the objects in the field with EW$_{[O III]} > 250$ \AA $ $ are C III] emitters.  While no such verification exists for the HDFS sample, our results \cite[as well as those of][]{2014MNRAS.445.3200S,2017arXiv170104416A} demonstrate that most if not all C III] emitters have strong optical emission lines that could contaminate broadband photometry and bias SED-derived results.  As such, the contamination of the near-IR $J$-, $H$-, and $K$-band photometry at these redshifts would bias us toward measuring stellar masses and ages that are too large by potentially 0.5 dex \citep{2009AA...502..423S}.  In practice, this contamination is offset by the longer wavelength \textit{Spitzer}/IRAC photometry, which is not contaminated by emission lines at these redshifts.  If we were systematically biased due to overestimating the continuum level, such a bias would change the offsets shown in Figure \ref{fig:exp} since we expect the changes in mass and age to be the largest for the objects with strong optical lines, namely the C III] emitters.  This would provide further evidence that the C III]-emitting population is younger than the nonemitting population and potentially demonstrate that such a discrepancy also exists for stellar mass.  To some extent these uncertainties are incorporated in the \magphys probability distributions, which we use in the histograms shown in Figure \ref{fig:exp} and in the determinations of the uncertainties in these parameters.  In future work we will incorporate corrections for the contamination by optical lines to the SED fitting, specifically focusing on data sets such as the UDF wherein existing near-IR spectroscopy can constrain their contribution.

Additionally, some of these discrepancies can be explained by selection effects.  For example, we are more likely to observe C III] in galaxies with low values of $A_V$ since extinction also lowers the flux of UV emission lines.  

\begin{figure*}
\begin{center}
\includegraphics[width=.9\textwidth]{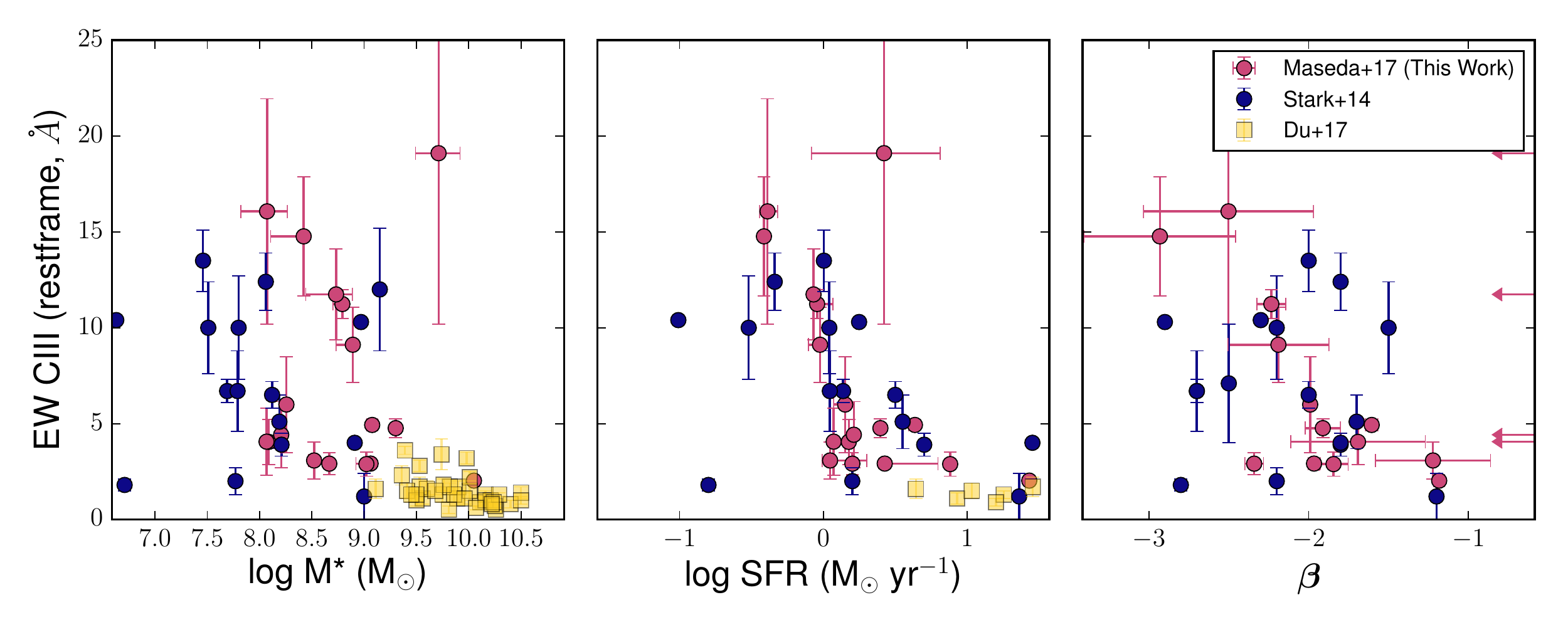}
\end{center}
\caption{C III] doublet EW versus SED-derived parameters (median and $\pm$ 1-$\sigma$ uncertainties) with the MUSE sample shown by pink points, galaxies from \citet{2014MNRAS.445.3200S} shown by blue points, and galaxies from \citet{2016arXiv161206866D} shown by yellow squares. HDFS galaxies with no constraints on $\beta$ are shown with arrows. \citet{2014MNRAS.445.3200S} SFRs come from their quoted sSFR values and their quoted M$_{\star}$ values.   In the case of stellar mass (SFR), EWs in excess of 5 \AA$~$are only found in galaxies with M$_{\star} \lesssim $ 10$^{9.5}$ \msol$~$($SFR \lesssim$ 10 \msol$~$yr$^{-1}$).}
\label{fig:ciiiparams}
\end{figure*}

We plot the values of C III] EW versus the derived values of stellar mass, star formation rate, and UV continuum slope for the galaxies in Figure \ref{fig:ciiiparams}.  When including the \citet{2014MNRAS.445.3200S} and \citet{2016arXiv161206866D} data points, there are no convincing trends between C III] EW and these parameters.  The \citet{2014MNRAS.445.3200S} galaxies are less massive on average, but this is expected given the boosting of fluxes through gravitational lensing.  It appears that, like metallicity, low stellar mass is a necessary but insufficient condition for high-EW C III] emission.  The relationship between SFR and EW is harder to interpret given that we are only able to observe C III] at high EW in galaxies that have low SFRs (and hence low UV luminosities).  Given the lack of clear \textit{trends}, we claim that it is not possible to predict the strength of C III] emission in galaxies based on SED-derived quantities such as stellar mass or star formation rate in contrast to the strong observed relation between [O III] emission and C III] emission in Figure \ref{fig:oiiiciii}.

\begin{figure}
\begin{center}
\includegraphics[width=.48\textwidth]{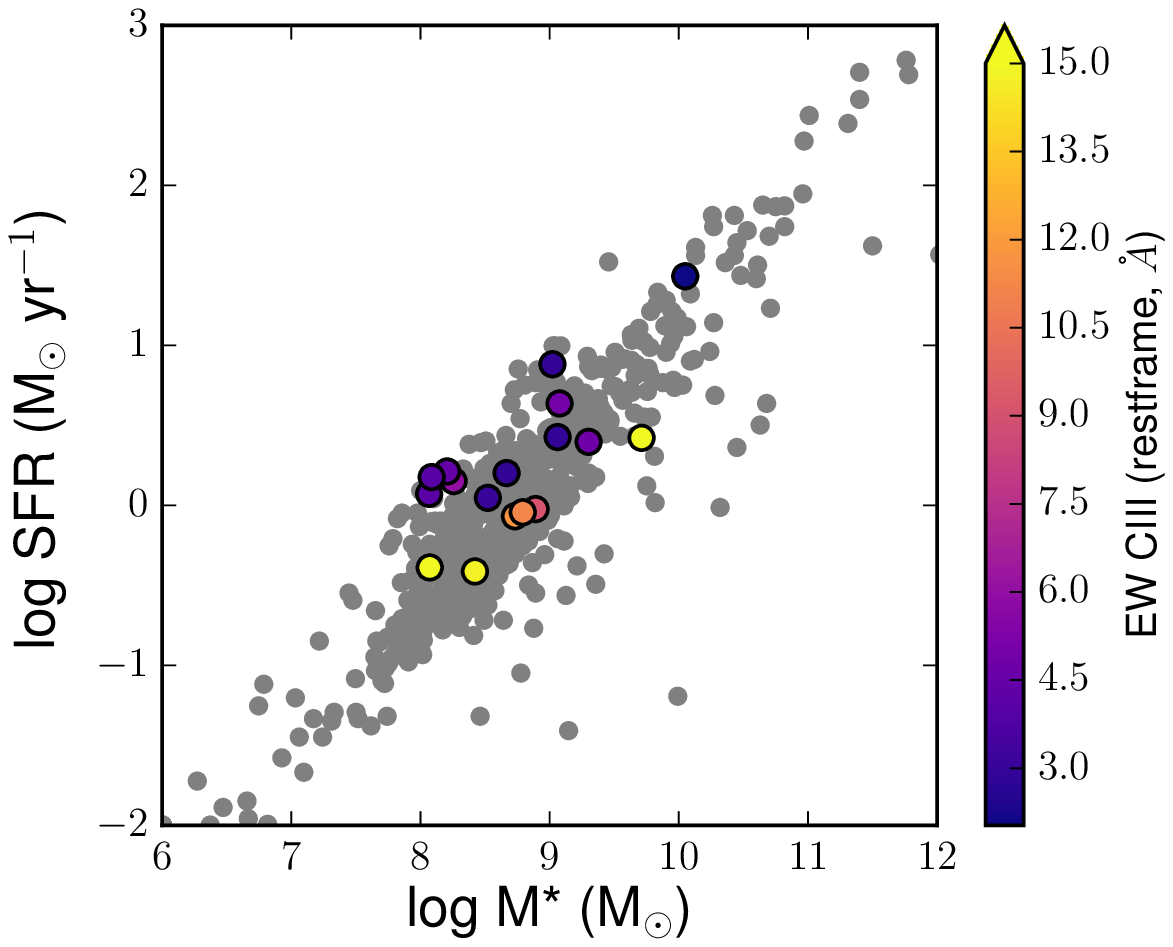}
%Dropbox/MUSE/magphys_pdf_stack_new
\end{center}
\caption{Stellar mass versus star formation rate for the median values from the \magphys probability distributions using the HDF-S and 3D-HST photometry.  C III] emitters are color-coded according to their rest-frame EW.  6 out of 318 galaxies (1.9\%) with $10^8$ \msol $~\leq M_{\star} \leq 10^9$ \msol$ $ and 1 out of 140 galaxies (0.7\%) with $10^9$ \msol $~\leq M_{\star} \leq 10^{10}$ \msol$ $ in this redshift range have a combined C III] EW in excess of 5 \AA.}
\label{fig:sfms}
\end{figure}

In Figure \ref{fig:sfms} we show the SED-derived stellar masses versus the SED-derived star formation rates with C III] emitters color coded by their C III] EW.  As shown in Figure \ref{fig:exp}, C III] emitters have average sSFR values that are higher than ``normal'' star-forming galaxies at these redshifts, but in Figure \ref{fig:sfms} we see that some of the C III] emitters lie in the same region of SFR-stellar mass space as the bulk of the galaxy population.  Using this information, we can constrain the fraction of high-EW C III] emitters in bins of stellar mass.  Six out of 318 galaxies (1.9\%) with $10^8$ \msol $~\leq M_{\star} \leq 10^9$ \msol$ $ and 1 out of 140 galaxies (0.7\%) with $10^9$ \msol $~\leq M_{\star} \leq 10^{10}$ \msol$ $ have a combined C III] EW in excess of 5 \AA.  Since we are flux limited in detecting C III], this must be considered a lower limit to the actual fraction of high-EW C III] emitters in these mass bins.  \citet{2016arXiv161206866D} do not find any C III] emitters with EWs in excess of 5 \AA$~$at z$\sim$1.  This makes sense in the context of Figure \ref{fig:ciiiparams}, where low stellar mass is a necessary condition for high-EW C III]; the median stellar mass of their sample is $10^{9.93}$ \msol.

\begin{figure}
\begin{center}
\includegraphics[width=0.48\textwidth]{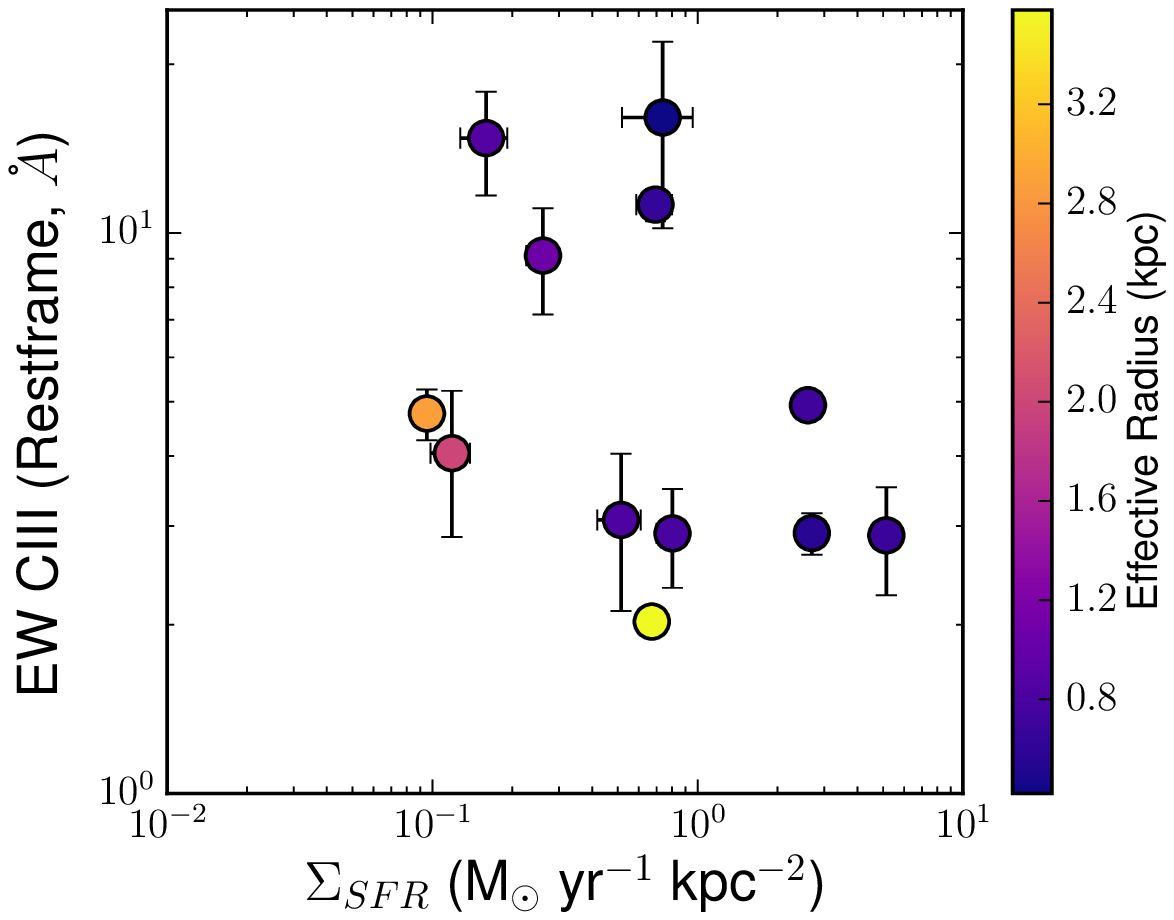}
%Dropbox/MUSE/magphys_ciii_plots.py -> on laptop
\end{center}
\caption{Observed star formation rate surface density, $\Sigma_{SFR}$, measured using the effective radius in kiloparsecs as measured in $J_{F125W}$ \citep{2014ApJ...788...28V} (color coding) vs. C III] EW for galaxies in the \udft.  At these redshifts, the $J_{F125W}$ band covers the rest-frame optical region of the galaxy SED probing the stellar continuum, whereas the C III] comes from the UV-dominated star-forming regions.  The HDFS region is not covered with HST imaging at these wavelengths.  The results are consistent when using the $H_{F160W}$ sizes.  Only the two largest galaxies here are resolved at the spatial resolution with MUSE.}
\label{fig:size}
\end{figure}

In Figure \ref{fig:size} we compare the rest-frame optical star formation surface density, measured using the effective radius \cite[measured in the $J_{F125W}$  band;][]{2014ApJ...788...28V} with the measured C III] EW for objects in the \udftns.  The HDFS region is not covered at near-infrared wavelengths with high spatial resolution HST imaging.  However, there is not a clear relation between $\Sigma_{SFR}$ and the EW of C III].  C III] is expected to arise in the dense star-forming regions that dominate the UV flux output of the young, star-forming galaxies, while the rest-frame optical sizes are indicative of the main stellar continuum.  All of our emitters with EW values in excess of 5 \AA $ $ have effective radii $\lesssim$ 1 kpc, which are similar to the systems presented in \citet{2017arXiv170104416A}, but this does not manifest as a higher $\Sigma_{SFR}$.  This is further evidence that the relation between C III] EW and global galaxy properties is complex.

The compact nature of the majority of C III] emitters at HST resolution vindicates our flux comparisons in Figure \ref{fig:diagnostic}, since the majority of the objects are essentially unresolved in the spectra at both the MUSE and HST/WFC3 resolution.

\section{Conclusions}
Using two 1$'~\times~$1$'$ MUSE data cubes with on-sky integration times $\sim$ 30 hours each, we construct a sample of 17 galaxies showing C III] $\lambda\lambda$1907,1909 in emission with equivalent widths in excess of 1 \AA $ $ at $1.5 < z < 3.9$.  Deep HST/WFC3 G141 near-infrared slitless grism spectroscopy in the area of the \udft cube shows that nearly all of the C III] emitters show bright, high-equivalent width optical emission lines such as [O II] $\lambda\lambda$3727,3729, H$\beta$, and [O III] $\lambda\lambda$4959,5007.  The equivalent width of [O III] increases with the equivalent width of C III], and every galaxy with an [O III] equivalent width in excess of 250 \AA $ $ (Extreme Emission Line Galaxies) in the \udft footprint has a detection of C III] with MUSE.  Photoionization models of star formation cannot fully reproduce our observed flux ratios, implying that the models may need further tuning of, for example, the C/O ratio, which can only be performed sensibly with larger samples.  Likewise, constraining the contribution of AGN and shocks to the ionization state of the galaxies can best be performed with larger samples that include other rest-frame UV emission features. We perform detailed fits of the SEDs of all galaxies in these fields using
broadband photometry that spans from ultraviolet to infrared wavelengths,.  This reveals that C III] emitters have different properties from the general population, namely lower dust obscuration, higher (s)SFR values, younger ages, and bluer continuum slopes.  Meaningful trends between C III] EW and physical parameters are difficult to establish, but in general high-EW C III] (> 5 \AA) only occurs at masses $\lesssim$ 10$^{9.5}$ \msol$~$and SFRs $\lesssim$ 10 \msol$~$yr$^{-1}$ (Figure \ref{fig:ciiiparams}), and at metallicities $\lesssim$ 0.5 $Z_{\odot}$.  Electron densities measured from the ratio of the 1907 to 1909 \AA $ $ component are similar to the values derived from local galaxies using [Ar IV], which have a similar ionization potential, and therefore we cannot conclusively claim that the densities of the interstellar medium are significantly different in C III] emitting galaxies at $z > 1.5$ than they are at $z\sim0$ star-forming galaxies.

These findings have many implications for studies of galaxies at the highest redshifts.  Until recently, the vast majority of redshift confirmations at $z\gtrsim4$ have come from detections of the Lyman-$\alpha$ emission line of hydrogen at a rest-frame wavelength of 1216 \AA.  Beyond $z\sim6$, the intergalactic medium is believed to be predominantly neutral and hence the emitted Lyman-$\alpha$ photons from galaxies would be absorbed.  C III] photons cannot ionize neutral hydrogen and are therefore unaffected by the neutral intergalactic medium.  C III] is also one of the brightest emission features in the rest-frame ultraviolet of young star-forming galaxies and the fact that it is a doublet aids in its unambiguous detection, so it is a promising feature to confirm the redshifts of galaxies at the earliest cosmic times.  However, these results show that at $1.5 < z < 4,$ strong C III] emission is still relatively rare with detections in only $\sim$3\% of galaxies.  This must be considered a lower limit to the prevalence of C III] at these redshifts, given that we detect two objects with line fluxes $<$ 10$^{-18}$ erg s$^{-1}$ cm$^{-2}$ where we are $\lesssim$ 5\% complete.  Based on the tight relationship between the equivalent widths of C III] and [O III] for the typical [O III] measured equivalent widths at z $\sim$ 6, we predict a C III] equivalent width of 4.3 \AA.  

In a photoionization sense, C III] and other rest-frame ultraviolet metal lines encode valuable information about the physical state of the gas in galaxies.  We can only properly constrain models to understand the early stages
of galaxy formation and evolution with observations at z $\sim1-4$, where we have easy access to other spectral lines at rest-frame ultraviolet and optical wavelengths.  While future instrumentation such as NIRSpec on the \textit{James Webb Space Telescope} with wavelength coverage from $\sim0.6-5.3~\mu$m extends the redshift range at which we are able to observe these features, we have the opportunity now to refine our models and provide crucial baselines for our understanding of the physical properties of galaxies.

The work presented here uses only a subset of the deep MUSE observations that are currently underway.  As presented in Paper I, there is a larger 3$'~\times~$3$'$ mosaic in the UDF to $\sim$10 hour depth, providing 4.5$\times$ the area of the \udft pointing with approximately half of the line flux sensitivity.  There is also a wider survey covering 100 square arcminutes in the COSMOS and GOODS-S CANDELS/3D-HST regions to 1 hour depth, a subset of which is presented in \citet{2017AA...606A..12H}.  These areas will be crucial to improve the number statistics of C III] emitters, particularly at the highest equivalent widths and in objects that show Lyman-$\alpha$ emission.  HST/WFC3 G102 and G141 grism data also exist in these regions to varying depths, which will also allow us to extend our comparisons between rest-frame UV and rest-frame optical emission lines and provide constraints on the contribution of nebular emission to the broadband photometry of the galaxies.  We also plan on incorporating other rest-frame ultraviolet emission lines into these analyses, such as Lyman-$\alpha$, C IV $\lambda\lambda$1548,1550, He II $\lambda$1640, O III] $\lambda\lambda$1661,1666, and [Si III] $\lambda\lambda$1883,1892, which, combined with sophisticated photoionization models, will allow for tight constraints on the physical state of the gas in these galaxies.  Additional information, such as velocity offsets and the spatial information provided by MUSE, will also constrain the dynamical state of the galaxies.

\begin{acknowledgements}
We would like to thank the anonymous referee for comments that improved the clarity and quality of the manuscript, Gabe Brammer and Elisabete da Cunha for help with their respective codes, and Nicolas Bouch\'e and David Carton for productive discussions.  This work has made extensive use of the \texttt{TOPCAT} \citep{2005ASPC..347...29T} and the \texttt{Astropy} \citep{2013AA...558A..33A} software packages.  JB was in part supported by FCT through Investigador FCT contract IF/01654/2014/CP1215/CT0003 and his work was in part supported by Funda\c{c}{\~a}o para a Ci{\^e}ncia e a Tecnologia (FCT) through national funds (UID/FIS/04434/2013) and by FEDER through COMPETE2020 (POCI-01-0145-FEDER-007672).  RB acknowledges support from the ERC advanced grant MUSICOS.  TC acknowledges support of the ANR FOGHAR (ANR-13-BS05-0010-02), the OCEVU Labex (ANR-11-LABX-0060) and the A*MIDEX project (ANR-11-IDEX-0001-02) funded by the ``Investissements d'avenir'' French government program managed by the ANR.  RAM acknowledges support by the Swiss National Science Foundation.  JR acknowledges support from the ERC starting grant CALENDS-336736.  LW acknowledges funding by the Competitive Fund of the Leibniz Association through grant SAW-2015-AIP-2.
\\
Funding for the SDSS and SDSS-II has been provided by the Alfred P. Sloan Foundation, the Participating Institutions, the National Science Foundation, the U.S. Department of Energy, the National Aeronautics and Space Administration, the Japanese Monbukagakusho, the Max Planck Society, and the Higher Education Funding Council for England. The SDSS website is http://www.sdss.org/.

The SDSS is managed by the Astrophysical Research Consortium for the Participating Institutions. The Participating Institutions are the American Museum of Natural History, Astrophysical Institute Potsdam, University of Basel, University of Cambridge, Case Western Reserve University, University of Chicago, Drexel University, Fermilab, the Institute for Advanced Study, the Japan Participation Group, Johns Hopkins University, the Joint Institute for Nuclear Astrophysics, the Kavli Institute for Particle Astrophysics and Cosmology, the Korean Scientist Group, the Chinese Academy of Sciences (LAMOST), Los Alamos National Laboratory, the Max-Planck-Institute for Astronomy (MPIA), the Max-Planck-Institute for Astrophysics (MPA), New Mexico State University, Ohio State University, University of Pittsburgh, University of Portsmouth, Princeton University, the United States Naval Observatory, and the University of Washington.

\end{acknowledgements}
%{\small \bibliography{paperslibrary.bib}}
\bibliographystyle{aa} % style aa.bst
\bibliography{muse}

\end{document}